\documentclass[12pt,a4paper]{article}

\usepackage[cp1251]{inputenc}
\usepackage[english]{babel}

\usepackage{indentfirst}
\usepackage{amsmath}
\usepackage{amssymb}
\usepackage[pdftex]{graphicx,color}

\begin{document}

\begin{flushright}
SI-HEP-2015-10 ~QFET-2015-11 
\end{flushright}
\bigskip



\begin{center}
\textbf{\LARGE  Hadronic effects and observables\\[2mm] 
in $B\to \pi\ell^+\ell^-$ decay at large recoil}

\vspace*{1.6cm}

{\large Christian~Hambrock, Alexander~Khodjamirian and Aleksey~Rusov\!
\footnote{on leave from the Department of Theoretical Physics, 
Yaroslavl State University, Russia}}

\vspace*{0.4cm}

\textsl{%
Theoretische Physik 1,
Naturwissenschaftlich-Technische Fakult\"at,\\
Universit\"at Siegen, D-57068 Siegen, Germany}
\vspace*{0.8cm}

\textbf{Abstract}\\[10pt]
\parbox[t]{0.9\textwidth}{
We calculate the amplitude of the rare flavour-changing neutral-current 
decay $B\to \pi\ell^+\ell^-$  at large recoil of the pion.
The nonlocal contributions in which the weak effective operators
are combined with the electromagnetic lepton-pair emission 
are systematically taken into account. These amplitudes 
are calculated at off-shell values of the lepton-pair mass squared, $q^2<0$, 
employing the operator-product expansion, 
QCD factorization and light-cone sum rules. 
The results are fitted to hadronic dispersion relations in $q^2$, 
including the intermediate vector meson contributions.
The dispersion relations are then used in the physical region
$q^2>0$. Our main result is  
the process-dependent addition  $\Delta C^{(B\pi)}_9(q^2)$  
to the Wilson coefficient $C_9$ obtained at  $4m_\ell^2<q^2\lesssim m_{J/\psi}^2$.
Together with  the $B\to \pi$ form factors from 
light-cone sum rules, this quantity
is used  to predict the differential rate, direct CP-asymmetry and isospin asymmetry
in $B\to \pi\ell^+\ell^-$. We also estimate the 
total rate of the rare decay $B\to \pi\nu\bar{\nu}$. }

\end{center}

\vspace*{3cm}

\newpage

\newpage
\section{Introduction}
\label{sect:intro}
The first measurement of the $B^+\to \pi^+\mu^+\mu^-$ decay 
by the LHCb Collaboration \cite{LHCb:Bpimumu} paved the way for 
more detailed measurements  of $b\to d \ell^+\ell^-$  
decays. These results will complement the
available data on  $b\to s\ell^+\ell^-$ decays,
providing new important insight in  
the dynamics of flavour-changing neutral-current (FCNC) 
transitions in Standard Model (SM)  and beyond.

One important feature of  exclusive $b\to d \ell^+\ell^-$ decays
is a non-vanishing direct CP-asymmetry. 
In Standard Model (SM) this effect is
caused by the interference between the dominant short-distance contributions 
of semileptonic and magnetic dipole operators 
and the contributions of other effective operators accompanied  
by the  electromagnetic lepton-pair emission.
The amplitudes of the latter contributions  
are process-dependent, and are defined as hadronic matrix elements 
of {\em nonlocal} operator products.  
Importantly, the parts of the exclusive $b\to d \ell^+\ell^-$
decay amplitudes proportional 
to $\lambda_u\equiv V_{ub}V^*_{ud}$ and $\lambda_c\equiv V_{cb}V^*_{cd}$ are
of the same order of Cabibbo suppression and, in addition to a relative  CKM phase, 
have different strong phases originating from the nonlocal amplitudes. 
The main goal of this work is to calculate the 
hadronic matrix elements of nonlocal contributions to $B\to \pi\ell^+\ell^-$
at large recoil of the pion, that is, at small and intermediate lepton-pair mass,
$q^2\ll m_B^2$ . 

An advanced  theoretical description of the exclusive semileptonic FCNC decays
was developed on the basis of  QCD factorization (QCDF) approach \cite{BBNS},
applied first to the decays $B\to K^{(*)} \ell^+\ell^-$ in Ref.~\cite{BFS}
and  to $B\to \rho \ell^+\ell^-$  in Ref.~\cite{BFS05}; see also
further applications to $B\to K\ell^+\ell^-$ \cite{BuchallaetalBKll}, 
and to $B\to \pi\ell^+\ell^-$ \cite{HKX}. 
An approach combining QCD light-cone sum rules with QCDF at $q^2>0$ for 
$B\to  K^{(*)} \ell^+\ell^-$ was used in \cite{ZL}. 

In QCDF, the nonlocal effects in these decays are described 
in terms of hard-scattering quark-gluon amplitudes 
with virtual photon emission, convoluted with 
light-cone distribution amplitudes (DAs)
of the initial $B$ meson and final light meson. 
Soft gluons, responsible for 
the onset of long-distance effects in the channel of the electromagnetic 
current, including vector resonance formation and nonfactorizable 
interactions  with initial and final meson remain 
beyond the reach of QCDF. Hence, these 
contributions have to be kept small, protected by their power suppression.
In particular, one avoids the intervals of 
lepton-pair mass squared $q^2$
in the vicinity of vector meson masses, $q^2\sim m_V^2$  ($V=\rho,\omega,...,J/\psi,..$),
where nonlocal effects are largely influenced by long-distance
quark-gluon dynamics.
This constraint defines the region of applicability of QCDF, that  is, 
roughly from $q^2_{\rm min}= 2 ~\mbox{GeV}^2$ up to  
$q^2_{\rm max}=6 ~\mbox{GeV}^2$. In this region, 
quark-hadron duality approximation 
is tacitly assumed for the contributions of 
radially excited and continuum hadronic  
states  with the quantum numbers of light vector mesons.

Note that in the $B\to \pi \ell^+\ell^-$ decay, as compared to
$B\to K \ell^+\ell^-$, the role of nonlocal effects 
related to $\rho$ and $\omega$  resonances in the $q^2$ channel  
grows due to the  current-current 
operators with large Wilson coefficients in the $\sim \lambda_u$ part. 
For the same reason, the weak annihilation 
combined with virtual photon emission, being suppressed  
in $B\to K \ell^+\ell^-$ decays, becomes 
one of the dominant nonlocal effects   in $B\to \pi \ell^+\ell^-$.
In the QCDF approach one describes the weak annihilation 
contribution \cite{BFS} in terms 
of a virtual photon emission off the spectator antiquark in the $B$ meson, 
followed  by the subsequent annihilation to a final pion state. 
The accuracy of this leading-power diagram approximation, 
presumably quite sufficient
for $B\to K \ell^+\ell^-$ decay,  becomes crucial for $B\to \pi \ell^+\ell^-$. 

In this paper we calculate 
the nonlocal effects in $B\to \pi\ell^+\ell^-$,
using the method formulated in Ref.~\cite{KMPW_charmloop} and  
applied in Ref.~\cite{KMW_BKll} to $B\to K \ell^+\ell^-$. 
One avoids  applying  QCDF  directly in the physical region  
$q^2>0$  and calculates the amplitudes of nonlocal contributions 
at deep spacelike $q^2<0$, $|q^2|\gg \Lambda_{QCD}^2$, 
where  the operator-product expansion (OPE)  
and QCDF can safely be used. The OPE contributions 
include the leading-order (LO) 
loops and weak annihilation,  the NLO perturbative  
corrections to the loops and hard spectator scattering.
Furthermore, we include important nonfactorizable 
soft-gluon effects via dedicated LCSR calculations of hadronic matrix elements. 
The amplitudes of nonlocal effects are 
then represented in a form of hadronic dispersion 
relations in the variable $q^2$ where vector mesons are included
explicitly. The residues of vector-meson poles related to nonleptonic 
$B\to V\pi$ decays are fixed, using experimental data 
and/or QCDF estimates. The nonresonant part of the 
hadronic dispersion integral is parametrized combining 
quark-hadron duality with a polynomial ansatz. 
Finally, the unknown parameters in the  dispersion relation, most 
importantly, the strong phases of resonance and nonresonant contributions 
are fitted to the  QCD calculation at $q^2<0$. 
The advantage of describing nonlocal contributions to the 
FCNC decay amplitude in terms of hadronic 
dispersion relation is that the latter is valid 
in the whole large-recoil region specified as $4m_\ell^2<q^2<m_{J/\psi}^2$.

In $B\to \pi \ell^+\ell^-$ decays the combinations of CKM factors $\lambda_{u}$ 
and $\lambda_{c}$ are comparable in size. Correspondingly, 
we have to calculate separately two hadronic matrix elements 
of nonlocal effects multiplying  $\lambda_{u}$ and $\lambda_{c}$. 
A similar CKM separation 
has to be done in the amplitudes of nonleptonic $B\to V \pi$ decays,
used  to fix the residues 
of vector-meson poles in the hadronic dispersion relations.
To obtain the separate parts of nonleptonic $B\to V\pi$ amplitudes 
for $V=\rho, \omega $  we employ the QCDF results \cite{BN03} 
and control the resulting amplitudes with the data on branching ratios 
and CP-asymmetries of these nonleptonic decays.

The plan of this paper is as follows. In Sec.~\ref{sec:nonloc} we 
present the structure of the $B\to \pi\ell^+\ell^-$ decay amplitude 
and define the hadronic matrix element of nonlocal contributions.
Sec.~\ref{sec:deltaC} contains a detailed calculation of these
amplitudes at $q^2<0$. In Sec.~\ref{sec:numerics} 
we perform the relevant numerical analysis. In Sec.~\ref{sec:nonlept}  
the necessary inputs for  the nonleptonic $B\to V \pi$
decay amplitudes are presented. Sec.~\ref{sec:disp} is devoted to the 
analysis  of the hadronic  dispersion relations. Matching the latter 
to the  result of QCD calculation, we then obtain  
$\Delta C^{(B\pi)}_9(q^2>0)$. 
In Sec.~\ref{sec:obs} our predictions 
for the observables  in the $B\to \pi\ell^+\ell^-$ decay are presented, 
including the decay rate, direct $CP$-asymmetry and isospin-asymmetry. 
In Sec.~\ref{sec:NP} we estimate the rate for 
$B\to \pi \nu \bar{\nu}$ decay and Sec.~\ref{sec:concl}.
contains the concluding discussion. The two appendices 
contain: (A) the operators and Wilson coefficients of the effective 
Hamiltonian of $b\to d \ell^+\ell^-$ transitions and 
(B) the QCDF expressions used for  the 
amplitudes of $B\to \rho(\omega)\pi$ nonleptonic  decays.        

\section{The  $B \to \pi \ell^+ \ell^-$ decay amplitude}
\label{sec:nonloc}
The effective weak Hamiltonian of the $b \to d \ell^+ \ell^-$  transitions $(\ell = e, \mu, \tau)$ has the following form \cite{Heff,BBL_Heff} in the SM  
:
\begin{equation}
\label{eq:Heff}
H^{b \to d}_{\rm eff} = 
\frac{4 G_F}{\sqrt 2} \left(
\lambda_u \sum\limits_{i=1}^{2} C_i \, {\cal O}_i^{u} 
+\lambda_c \sum\limits_{i=1}^{2} C_i \, {\cal O}_i^{c}
-\lambda_t \sum\limits_{i=3}^{10} C_i \, {\cal O}_i 
\right) + h.c.\,,
\end{equation}
where 
$\lambda_p = V_{pb} V_{pd}^*$, $(p = u,c,t)$
are the products of CKM matrix elements.
In contrast to the $b \to s \ell^+\ell^-$ transitions, all three terms in 
the unitary relation have  the same order of Cabibbo suppression, 
$\lambda_u \sim \lambda_c \sim \lambda_t \sim \lambda^3$, $\lambda$ being
the Wolfenstein parameter. 
Hereafter, we assume CKM unitarity and replace 
$\lambda_t=-(\lambda_u + \lambda_c)$.
The local  dimension-6 operators  ${\cal O}_i$ in (\ref{eq:Heff}) 
together with the numerical values of their Wilson coefficients
$C_i$ at relevant scales are presented in the Appendix~A. 
\begin{figure}[t]\center
\label{fig:diagO7910}
\includegraphics[scale=0.6]{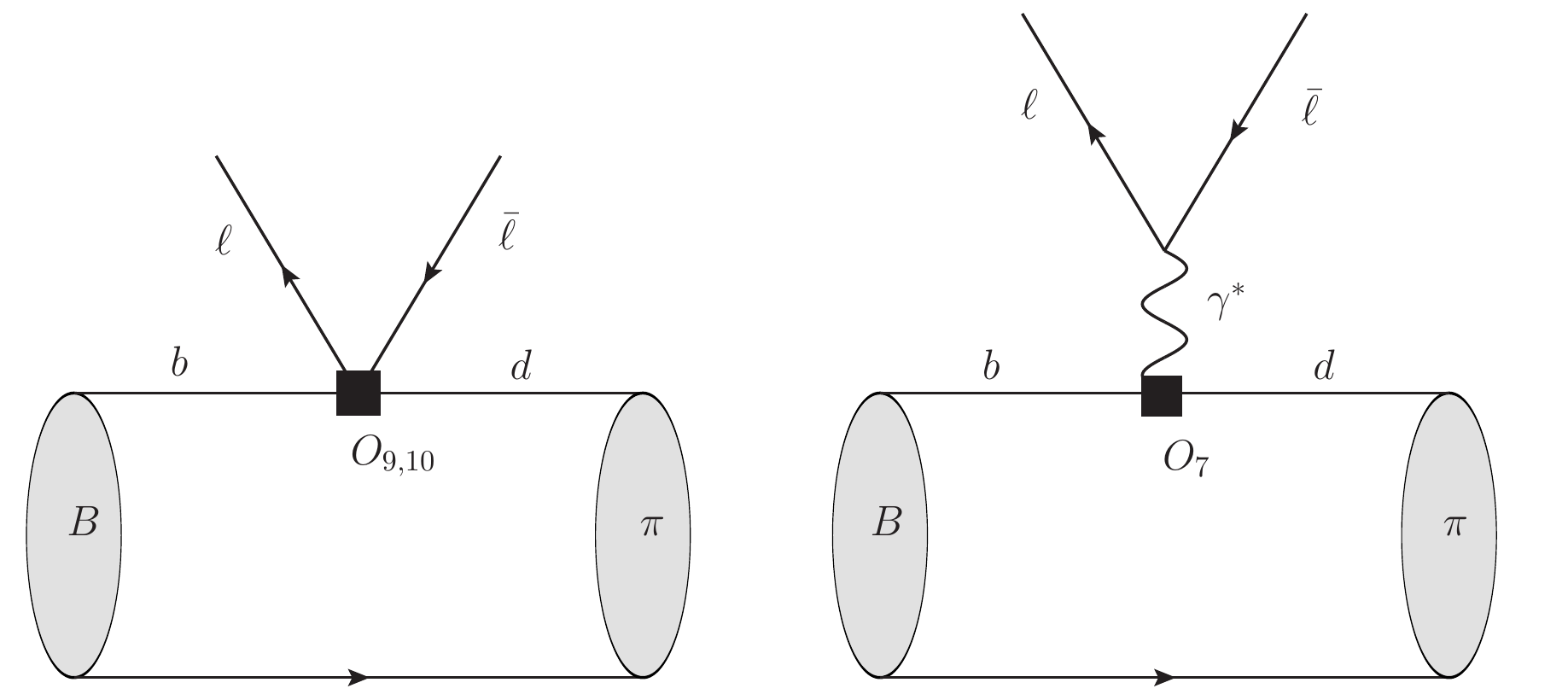}
\caption{\it FCNC contributions to $B \to \pi \ell^+ \ell^-$ due to the effective operators
$O_{9,10}$ (left) and $O_{7\gamma}$ (right) denoted as black squares.}
\end{figure}

The amplitude of the $B \to \pi \ell^+ \ell^-$ decay reads:
\begin{eqnarray}
A(B \to \pi \ell^+ \ell^-) & = & 
- \langle \pi (p) \ell^+ \ell^- | H_{\rm eff}^{b \to d } | B(p+q) \rangle 
\nonumber  \\
& = & \frac{G_F}{\sqrt 2} \frac{\alpha_{\rm em}}{\pi} \lambda_t
\Biggl[ \left(\bar \ell \gamma^\mu \ell \right) p_\mu \left(C_9 f^+_{B\pi} (q^2) 
+ \frac{2 m_b}{m_B + m_\pi} C_7^{\rm eff} f^T_{B\pi} (q^2) \right) \nonumber \\
 & + & \left(\bar \ell \gamma^\mu \gamma_5 \ell \right) p_\mu C_{10} f_{B\pi}^+ (q^2) 
+ 16 \pi^2 \frac{\bar \ell \gamma^\mu \ell}{q^2} \left( \frac{\lambda_u}{\lambda_t} {\cal H}_\mu^{(u)} + \frac{\lambda_c}{\lambda_t} {\cal H}_\mu^{(c)} \right) \Biggr],
\label{eq:ampl}
\end{eqnarray}
where $p^\mu$ and $q^\mu$ are the four-momenta of the 
$\pi$-meson and lepton pair, respectively. 
 
In (\ref{eq:ampl}), 
the dominant contributions of the  
operators $O_{9,10}$ and $O_{7\gamma}$ are separated 
(Fig. \ref{fig:diagO7910}) and their hadronic matrix elements are 
expressed in terms of the vector and tensor $B\to \pi$ form factors,
$f^+_{B\pi} (q^2)$ and $f^T_{B\pi} (q^2)$, respectively, 
defined in the standard way:
\begin{eqnarray}
\label{eq:fpl}
\langle \pi (p) | \bar d \gamma^\mu b | B (p+q) \rangle =  
f^+_{B\pi}  (q^2) \left [2 p^\mu + \left(1 - \frac{m_B^2 - m_\pi^2}{q^2} \right)q^\mu \right] 
\nonumber\\
+ f^0_{B\pi}  (q^2) \, \frac{m_B^2 - m_\pi^2}{q^2} \, q^\mu,  
\end{eqnarray}
\begin{equation}
\langle \pi (p) | \bar d \sigma^{\mu\nu} q_\nu b | B (p + q) \rangle =  \frac{i f^T_{B\pi}  (q^2)}{m_B + m_\pi} 
\left[ 2q^2 p^\mu + \bigg(q^2 - \left ( m_B^2 - m_\pi^2 \right )\bigg) q^\mu \right] \,. 
\end{equation}
For definiteness, 
hereafter we consider the $B^-\to \pi^-\ell^+\ell^-$ mode,
unless stated otherwise.
We assume isospin symmetry
for the $b\to d$   and $b\to u$ transition form factors. The $B^-\to \pi^-$ 
form factor $f^{+}_{B\pi}$ in Eq.~(\ref{eq:fpl})  is equal  to the one 
in the $\bar{B}^0\to \pi^+\ell^- \nu_\ell$
semileptonic decay  and the form factors in the   
$\bar{B}^0\to \pi^0\ell^+\ell^-$ decay amplitude have an extra factor 
$1/\sqrt{2}$.  For the $CP$-conjugated modes
$B^+\to \pi^+\ell^+\ell^-$ and $B^0\to \pi^0\ell^+\ell^-$,
respectively, one has to use the hermitian conjugated effective 
operators with complex conjugated CKM factors $\lambda_p^*$.

The  current-current, quark-penguin 
and chromomagnetic operators  in the effective Hamiltonian (\ref{eq:Heff})
contribute to the decay amplitude (\ref{eq:ampl}), with 
the lepton pair produced via virtual photon.
After factorizing out the lepton pair, the  expression 
for these nonlocal effects is arranged in (\ref{eq:ampl}) in a form of correlation functions
of the time-ordered product of quark operators with the quark
e.m. current,  
$j_\mu^{\rm em} = \sum_{q = u,d,s,c,b} Q_q \bar q \gamma_\mu q$,
sandwiched between $B$ and $\pi$ states:   
\begin{eqnarray}
\label{eq:corr}
{\cal H}_\mu^{(p)} & =&  i \int d^4 x e^{i q x} \langle \pi (p) |{\rm T} \biggl\{ j_\mu^{\rm em} (x), \biggl[C_1 {\cal O}_1^{p} (0) + C_2 {\cal O}_2^{p} (0) \nonumber \\
+ & & \hspace*{-10mm}\sum\limits_{k=3-6,8g} C_k {\cal O}_k (0) \biggr] \biggr\} | B (p+q) \rangle = 
\left[ (p \cdot q) q_\mu - q^2 p_\mu \right] {\cal H}^{(p)} (q^2),~~(p=u,c),
\end{eqnarray}
where the index $p=u,c$  hereafter distinguishes
the hadronic matrix elements in Eq.(\ref{eq:ampl}),
multiplying, respectively, the CKM factors $\lambda_u,\lambda_c$.  
Substituting (\ref{eq:corr}) in (\ref{eq:ampl}) and taking
into account the conservation of the leptonic current, we write down 
the decay amplitude in a more compact form:
\begin{eqnarray}
\label{eq:Bpill}
A(B \to \pi \ell^+ \ell^-) 
=  \frac{G_F}{\sqrt 2} \lambda_t\frac{\alpha_{\rm em}}{\pi} 
f^+_{B\pi}(q^2)\Biggl[ \left(\bar \ell \gamma^\mu \ell \right) p_\mu 
\Bigg(C_9 + \Delta C^{(B\pi)}_9(q^2)
\nonumber\\
+ \frac{2 m_b}{m_B + m_\pi} C_7^{\rm eff} r_{B\pi}^T (q^2) \Bigg)
 +  \left(\bar \ell \gamma^\mu \gamma_5 \ell \right) p_\mu 
C_{10} \Biggr]\,, 
\end{eqnarray}
where the invariant amplitudes introduced in Eq.~(\ref{eq:corr})
form a process-dependent and $q^2$-dependent 
addition to the Wilson coefficient $C_9$:
\begin{equation}
\label{eq:Delta-C9}
\Delta C_9^{(B\pi)} (q^2)  
 \equiv - 16\pi^2\frac{\left( \lambda_u {\cal H}^{(u)} (q^2) + 
\lambda_c {\cal H}^{(c)} (q^2) \right)}{\lambda_t f_{B\pi}^+ (q^2) }\,.
\end{equation}
In Eq.(\ref{eq:Bpill}) we also introduce the ratio of tensor and vector form factors:
\begin{equation}
r_{B\pi}^T (q^2) \equiv \frac{f_{B\pi}^T (q^2)}{f_{B\pi}^+ (q^2)} \,.
\label{eq:rT}
\end{equation}
In the case of $b \to s \ell^+ \ell^-$ transitions, 
the factor $\lambda_u$ is usually neglected 
so that $\lambda_c = -\lambda_t$, and one recovers the 
corresponding expression for $\Delta C_9^{(BK)} (q^2)$ in $B\to K \ell^+\ell^-$
used  in Ref.~\cite{KMW_BKll}.

\section{Nonlocal effects at spacelike $q^2$}
\label{sec:deltaC}
In this section we present separate contributions to the nonlocal amplitudes 
${\cal H}^{(u)}(q^2)$  and ${\cal H}^{(c)}(q^2)$ defined in (\ref{eq:corr}) and 
calculated at $q^2 <0$, in the same approximation that was
adopted in Ref.~\cite{KMW_BKll} for $B \to K \ell^+ \ell^-$.

\subsection{Factorizable loops}
\begin{figure}\center
\includegraphics[scale=0.6]{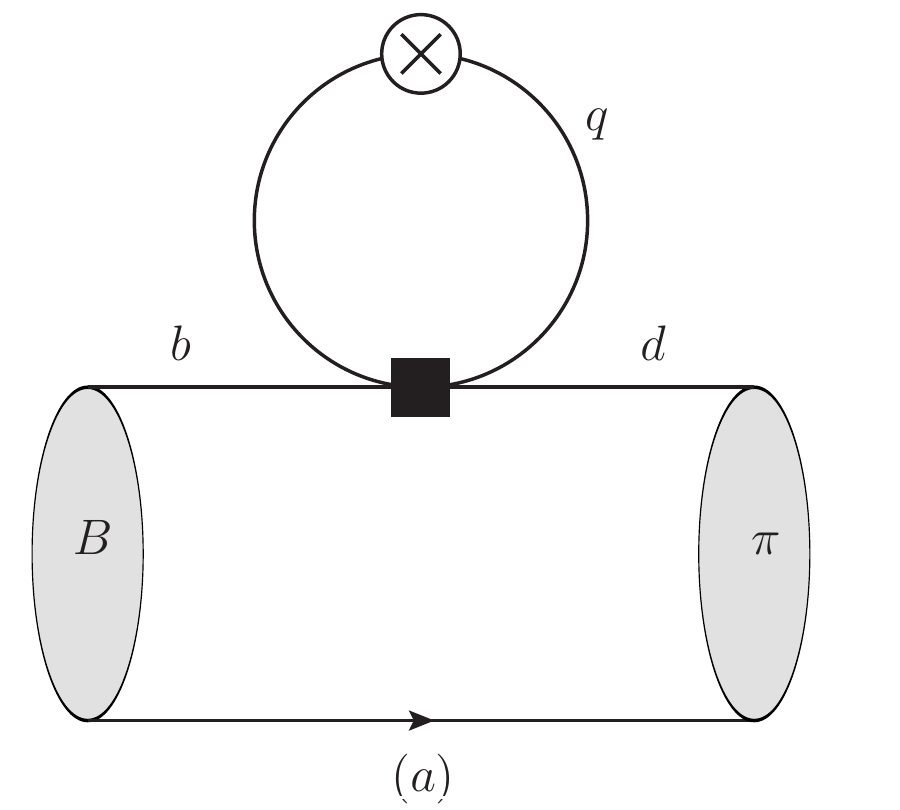}~~
\includegraphics[scale=0.4]{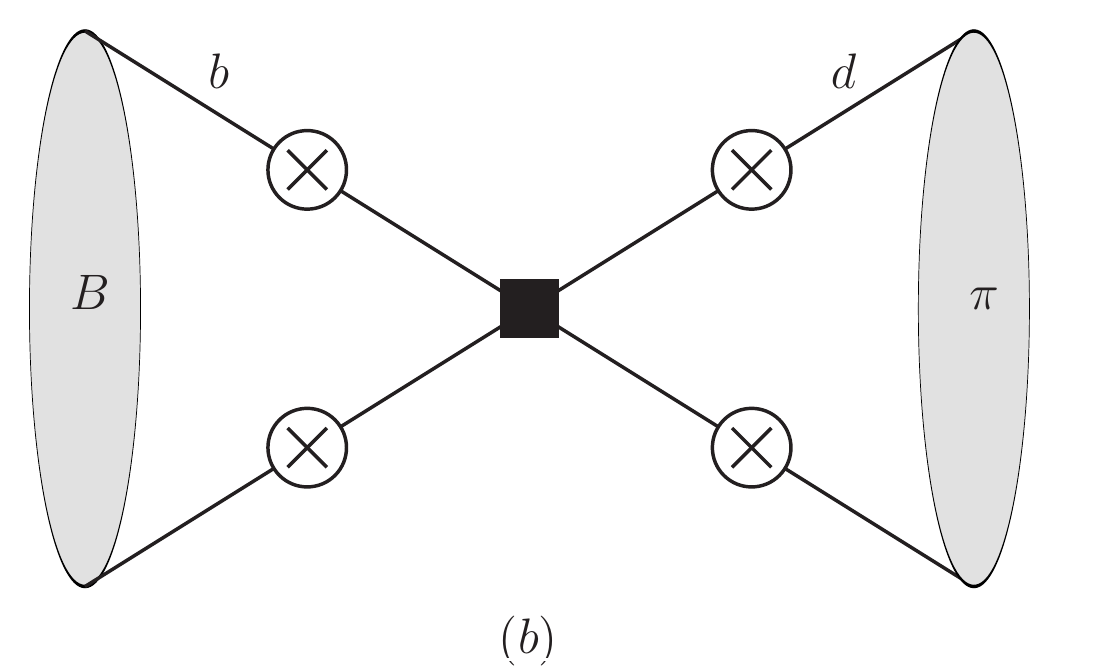}\\
\caption{ \it Leading-order diagrams of nonlocal effects  
in $B \to \pi \ell^+ \ell^-$
due to the four-quark effective operators $O_{1,2}^{u,c}$ and $O_{3-6}$: 
the quark-loop (a) and weak annihilation (b).
The black square denotes the operator
and crossed circles  indicate possible points of the virtual photon emission.}
\label{fig:LO-fact}
\end{figure}

At leading order (LO) in the quark-gluon coupling, the contributions of 
the four-quark operators to $B \to \pi \ell^+ \ell^-$
have two possible quark topologies. One of them corresponds to 
the factorizable quark-loop  diagrams with different flavours
(see Fig.~\ref{fig:LO-fact}(a)). 
Their expressions are obtained from, e.g., the ones 
presented in Ref.~\cite{KMW_BKll}, separating  the $\sim \lambda_u$ and $\sim \lambda_c$ parts: 
\begin{eqnarray}
{\cal H}^{(u)}_{\rm fact, LO}(q^2) & = & \frac{1}{12 \pi^2} 
\left(\frac{C_1}{3} + C_2 \right)g_0(q^2)f^+_{B \pi} (q^2)+
{\cal H}^{(3-6)}_{\rm fact, LO}(q^2)\,,
\label{eq:HuLO}\\
{\cal H}^{(c)}_{\rm fact, LO}(q^2) & = & \frac{1}{12 \pi^2} 
\left(\frac{C_1}{3} + C_2 \right)g(q^2, m_c^2)f^+_{B \pi} (q^2)+
{\cal H}^{(3-6)}_{\rm fact, LO}(q^2)\, ,
\label{eq:HcLO}
\end{eqnarray}
where the common term stemming from the 
quark-penguin operators $O_{3-6}$  is 
\begin{eqnarray}
{\cal H}^{(3-6)}_{\rm fact, LO}(q^2) & = & \frac{1}{24 \pi^2} 
\Biggr[ -  \left(\frac 4 3 C_3 + \frac 4 3 C_4 + C_5 + \frac{C_6}{3} \right) 
\left(g(q^2, m_b^2)+g(q^2, m_s^2) \right) \nonumber \\
 & + & 2 \left(C_3 + \frac 1 3 C_4 + C_5 + \frac{C_6}{3} \right) g(q^2, m_c^2) 
\nonumber \\
& 
+ 
&   
\left(C_3 + \frac{C_4}{3} + C_5 + \frac{C_6}{3} \right) g_0 (q^2) 
\nonumber \\
& 
+ 
&  
\left( C_3 + \frac{C_4}{3} + C_5 + \frac{C_6}{3} \right) \Biggl] f^+_{B \pi} (q^2)\,.
\end{eqnarray}
For the loop function we use the expression valid at $q^2 < 0$:
\begin{eqnarray}
g(q^2,m_q^2) & = & \frac{4 m_q^2}{q^2} + \frac 2 3 - \ln{\frac{m_q^2}{\mu^2}}  + \sqrt{1 - \frac{4 m_q^2} {q^2}} \left(\frac{2 m_q^2}{q^2} + 1 \right) \nonumber \\
& \times & \ln\left(\frac{\sqrt{4 m_q^2 - q^2} - \sqrt{-q^2}} {\sqrt{4 m_q^2 - q^2} + \sqrt{-q^2}} \right)\,,
\label{eq:mloop}
\end{eqnarray}
where $m_q$ is the quark mass if $q = b, c, s$ and $\mu$ 
is the renormalization scale. 
For $u$- and $d$-quark loops the quark masses are neglected; in this case 
the loop function takes the form: 
\begin{equation}
g_0 (q^2) = \lim\limits_{m_q^2 \to 0} g(q^2, m_q^2) = \frac 2 3 - \ln\left({\frac{-q^2}{\mu^2}}\right).
\label{eq:m0loop}
\end{equation}
In Eqs.~(\ref{eq:HuLO}) and (\ref{eq:HcLO}) the ``full'' $B\to \pi$ form factor
is the same as in the 
contributions of $O_{9,10}$ operators to the decay amplitude (\ref{eq:ampl}). 
For this form factor we will use LCSR results that are valid also at 
$q^2<0$. 

Note that in the LO approximation, 
when gluon exchanges between the loop and the rest of the diagram 
in Fig~\ref{fig:LO-fact}(a) are neglected, the nonlocal amplitudes
$H^{(u,c)}_{\rm fact, LO}(q^2)$ can also be calculated  within LCSR approach.
One has to define the vacuum-to-pion 3-point correlation function of the $B$-meson 
interpolating current, the four-quark operator and the 
electromagnetic current. After the quark loop  is factorized out
at large spacelike $q^2$,  the remaining correlation function
coincides with the one used to calculate the $B\to \pi$ form factor
from LCSR. The resulting sum rule is then reduced to the loop factor 
multiplied by the LCSR  expression for the $B\to \pi$ form factor,   
reproducing  Eqs.~(\ref{eq:HuLO}), (\ref{eq:HcLO}).   

\subsection{Weak annihilation}

The second possible topology at LO is the weak annihilation (WA) with 
the diagrams shown in Fig.~\ref{fig:LO-fact}(b).
In QCDF, neglecting the inverse heavy $b$-quark mass corrections, 
the leading diagram is the one where the virtual photon is emitted off the spectator quark 
$q=u,d$ in the $B$ meson,  with the resulting expression \cite{BFS,BuchallaetalBKll}: 
\begin{equation}
\label{WA-int-expr}
{\cal H}^{(p)}_{\rm WA}(q^2) = \frac{1}{8 N_c} \frac{f_B f_\pi m_b}{m_B^2}  \int\limits_0^\infty \frac{d\omega}{\omega} \phi_B^- (\omega) \int\limits_0^1 d u \, \varphi_\pi (u) \, T_-^{(0),p} (u,\omega), \quad (p = u, c), 
\end{equation}
where $f_\pi$ and $f_B$ are the $\pi$- and $B$-meson decay constants, respectively, 
and the hard-scattering amplitude
\begin{equation}
\label{eq:kernel}
T_-^{(0),p} (u, \omega) = Q_q\widetilde C_{\rm WA}^{p}  \frac{4 m_B}{m_b}
\frac{m_B \omega}{m_B \omega - q^2}\,, \quad (p=u,c)
\end{equation}
is convoluted with the $B$-meson DA  $\phi^-_B (\omega)$ 
defined as in Refs.~\cite{GN,BF} and  $\varphi_\pi (u)$ is the twist-2 pion DA.
The factor
\begin{equation}
\tilde C_{\rm WA}^{p} = 
\delta_{pu} (\delta_{qu} (C_2 + 3 C_1) + \delta_{qd} (C_1 + 3 C_2)) 
+C_3 + 3 C_4\,
\end{equation}
is the combination of Wilson coefficients depending on the flavour-content 
of the $B$ meson. 
To obtain the amplitudes 
${\cal H}^{(p)}_{\rm WA}(q^2)$,
one takes into account that the  LO  kernel (\ref{eq:kernel}) is 
independent of the variable $u$, hence the integral over  $\varphi_\pi (u)$ is reduced to its unit normalization.  Adopting the exponential ansatz \cite{GN} for the $B$-meson DAs:
\begin{equation}
\phi_B^+ (\omega)= \frac{\omega}{\lambda_B^2}e^{-\omega/\lambda_B},~~~
\phi_B^- (\omega)= \frac{1}{\lambda_B}e^{-\omega/\lambda_B}\,,
\label{eq:BDAansatz}
\end{equation}
where $\lambda_B$ is the inverse moment,
we obtain 
the following expression for the amplitude valid at $q^2 < 0$:
\begin{equation}
\label{WA-neg-qsq}
{\cal H}^{(p)}_{\rm  WA}(q^2) = - \frac{Q_q f_B f_\pi}{2 N_c m_B \lambda_B}
e^{-q^2/{m_B \lambda_B}} {\rm Ei} \left(\frac{q^2}{m_B \lambda_B}\right) 
\tilde C_{\rm WA}^{p}\,,
\end{equation}
where 
${\rm Ei} (x) = - \int\limits_{-x}^{\infty} d t e^{-t}/t$.

In contrast to $B\to K \ell^+ \ell^-$ transitions, the WA mechanism due to 
the enhanced current-current operators  $O_{1,2}^{u}$, 
provides  one of the dominant contribution to the 
${\cal H}^{(u)}_{\rm WA}(q^2)$ amplitude in 
$B\to \pi \ell^+\ell^-$. Moreover, the resulting difference 
between the WA amplitudes in 
$B^-\to \pi^-\ell^+ \ell^- $ and $\bar{B}^0\to \pi^0\ell^+ \ell^-$ 
contributes to the isospin asymmetry in $B\to \pi \ell^+\ell^-$.

Since the role of WA effects becomes important, it is 
desirable to improve the accuracy beyond the 
leading diagram contribution considered here. 
We checked that adding all subleading diagrams
 in Fig.~\ref{fig:LO-fact}(b) to the  virtual photon emission  
from  the spectator quark  does not produce a visible 
effect for the $O_{1,2}$ contributions.
There still remain power suppressed corrections generated by the 
higher twists in the pion DAs, and the contributions of 
the operators $O_{5,6}$ yet unaccounted in QCDF. In the future also  the 
perturbative nonfactorizable corrections to the diagrams in Fig.~\ref{fig:LO-fact}(b)
have to be calculated. 

In principle, it is also possible to calculate the WA
contribution employing the LCSR 
approach with the $B$-meson  DAs. The correlation function
will be described by the diagram similar to Fig.~\ref{fig:LO-fact}(b), but  
with the on-shell pion  replaced by the interpolating 
quark current  with the virtuality $p^2$. After employing the hadronic dispersion
relation and quark-hadron duality in the pion channel, 
in the factorizable approximation, 
the two-point part of this correlation function
will yield the QCD sum rule for the pion decay constant squared. 
The result for ${\cal H}^{(p)}_{\rm WA}(q^2)$ will then yield  
the expression (\ref{WA-neg-qsq}).

\subsection{Factorizable NLO contributions}

\begin{figure}\center
\includegraphics[scale=0.45]{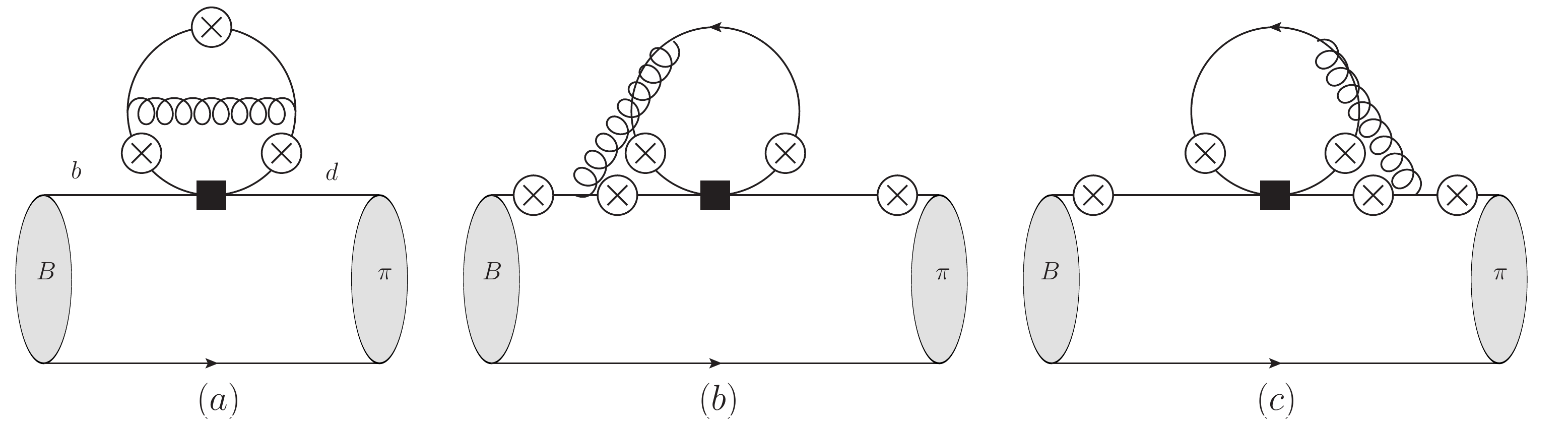}\\[2mm]
\includegraphics[scale=0.5]{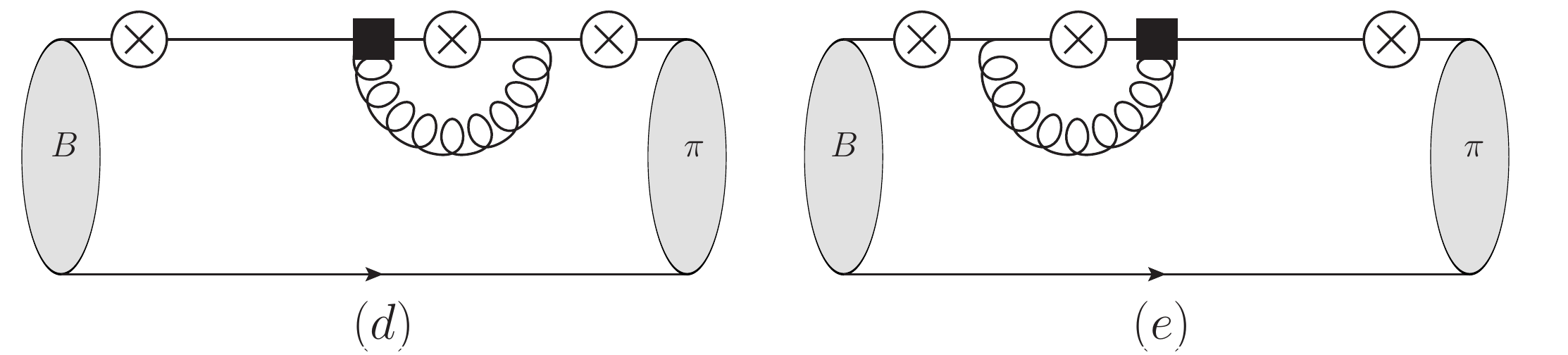}\\
\caption{\it Factorizable NLO quark-loop contributions to $B \to \pi \ell^+ \ell^-$, due to the four-quark effective operators $O_{1,2}^{u,c}$ 
(upper panel) and the chromomagnetic operator $O_{8g}$ (lower panel). 
The notation is the same as in Fig.~2.}
\label{fig:NLO-fact}
\end{figure}

The NLO corrections to the quark loops generated by the current-current operators $O^{u,c}_{1,2}$ are given by the two-loop diagrams shown
in Fig.~\ref{fig:NLO-fact}\,(a),(b),(c). The contributions 
of quark-penguin operators are neglected, being suppressed by small
Wilson coefficients and $\alpha_s$ simultaneously. 
In the same order of the perturbative expansion, 
the chromomagnetic operator $O_{8g}$ is described with the diagrams 
shown in Fig.~\ref{fig:NLO-fact}(d),(e).
These factorizable NLO contributions were taken into account in QCDF 
\cite{BFS,BFS05}, employing the quark-level two-loop diagrams calculated in 
Ref.~\cite{Asatryan:2001zw} (see also Refs.~\cite{Asatrian:2003vq,Seidel:2004jh}). 
The NLO contribution of $O_{1,2}^{c}$  to ${\cal H}^{(c)}$
can be literally taken from Ref.~\cite{KMW_BKll}, replacing 
$f^+_{BK} (q^2)$ by $f^+_{B \pi} (q^2)$. The corresponding contribution 
to ${\cal H}^{(u)}$ has the same structure, so that we can present both contribution 
by one compact expression:
\begin{eqnarray}
{\cal H}^{(p)}_{\rm fact,NLO}  =  - \frac{\alpha_s}{32 \pi^3} \frac{m_b}{m_B}
\Biggl\{C_1 F_{2,p}^{(7)} (q^2) + C_8^{\rm eff} F_8^{(7)} (q^2) 
\nonumber \\
+ 
\frac{m_B}{2 m_b} \biggl[C_1 F_2^{(9)} 
+  2 C_2 \left(F_{1,p}^{(9)} (q^2) + \frac 1 6 F_{2,p}^{(9)} (q^2) \right) + 
C_8^{\rm eff} F_8^{(9)} (q^2) \biggr] \Biggr\} f^+_{B \pi} (q^2),
\end{eqnarray}
where $p=u,c$. The definitions and nomenclature of the indices of 
the functions $F^{(7,9)}_{1,p}$, $F^{(7,9)}_{2,p}$ and $F_8^{(7,9)}$  
are the same as  in Refs.~\cite{Asatryan:2001zw,Asatrian:2003vq}. 
The only difference is that $F_{1,c}^{(7,9)}$  and $F_{2,c}^{(7,9)}$ are expressed as a
double expansion in $\hat s = q^2/m_b^2$ and $\hat m_c^2 = m_c^2/m_b^2$, 
whereas $F_{1,u}^{(7,9)}$ and $F_{2,u}^{(7,9)}$ are expanded only 
in powers $\hat s = q^2/m_b^2$  since we work in the limit $m_u = 0$.
It has been shown in Refs.~\cite{Asatryan:2001zw,Asatrian:2003vq} that keeping 
the terms up to the third power of $\hat s$ and $\hat m_c^2$ 
provides a sufficient numerical accuracy in the 
region $0.05 \leq \hat s \leq 0.25$. Here we use this expansion for $q^2 <0$,
restricting ourselves to 
$1.0 \leq |q^2| \leq 4.0 $ GeV$^2$, i.e. staying within the same  region. 
For $F_8^{(7,9)}$ we use the expression derived in Ref.~\cite{BFS}.

We remind that at NLO, the
nonlocal contributions acquire the imaginary part 
also at $q^2<0$,  that is, not related to the singularities in the variable $q^2$.
The origin of this imaginary part and its relation to the final-state
strong interaction is the same as for $B\to K \ell^+\ell^-$  
and is explained  in detail  in Ref.~\cite{KMW_BKll}.  

Note that analytic expressions for the two-loop virtual 
corrections to the matrix elements of the $O_1^{u}$ and $O_2^{u}$ operators
are available from Ref.~\cite{Seidel:2004jh}. These expressions 
are valid at $q^2>0 $ and agree with the expansion in $q^2 /m_b^2$ 
obtained in Ref.~\cite{Asatrian:2003vq}. However, we cannot use the
results of Ref.~\cite{Seidel:2004jh} straightforwardly 
in our calculation at $q^2<0$, without separating the imaginary 
contributions inherent to the negative $q^2$-region
from the contributions appearing due to 
the cuts of quark-gluon diagrams at  $q^2>0$.
Hence, we prefer to use the expanded form of these corrections  
\cite{Asatrian:2003vq} in which the phases stemming from
the positivity of $q^2$,  e.g., the terms proportional to 
$i\pi$ and originating from the $\log q^2$  terms  
can be easily recognized and separated.
As we work at sufficiently small
values of $|q^2|$, the accuracy of the expansion in  Ref.~\cite{Asatrian:2003vq} 
is sufficient.

Contrary to the LO contributions considered in the previous subsections, 
the factorizable NLO ones are not simply accessible within the LCSR approach. 
Indeed, in order to reach the same  $O(\alpha_s)$ accuracy, 
the calculation of the underlying correlation function has to include  
two-loop  diagrams with several scales, a task exceeding 
the currently reached level of complexity  in the multiloop calculations.

\subsection{Nonfactorizable soft-gluon contributions}

\begin{figure}\center
\includegraphics[scale=0.5]{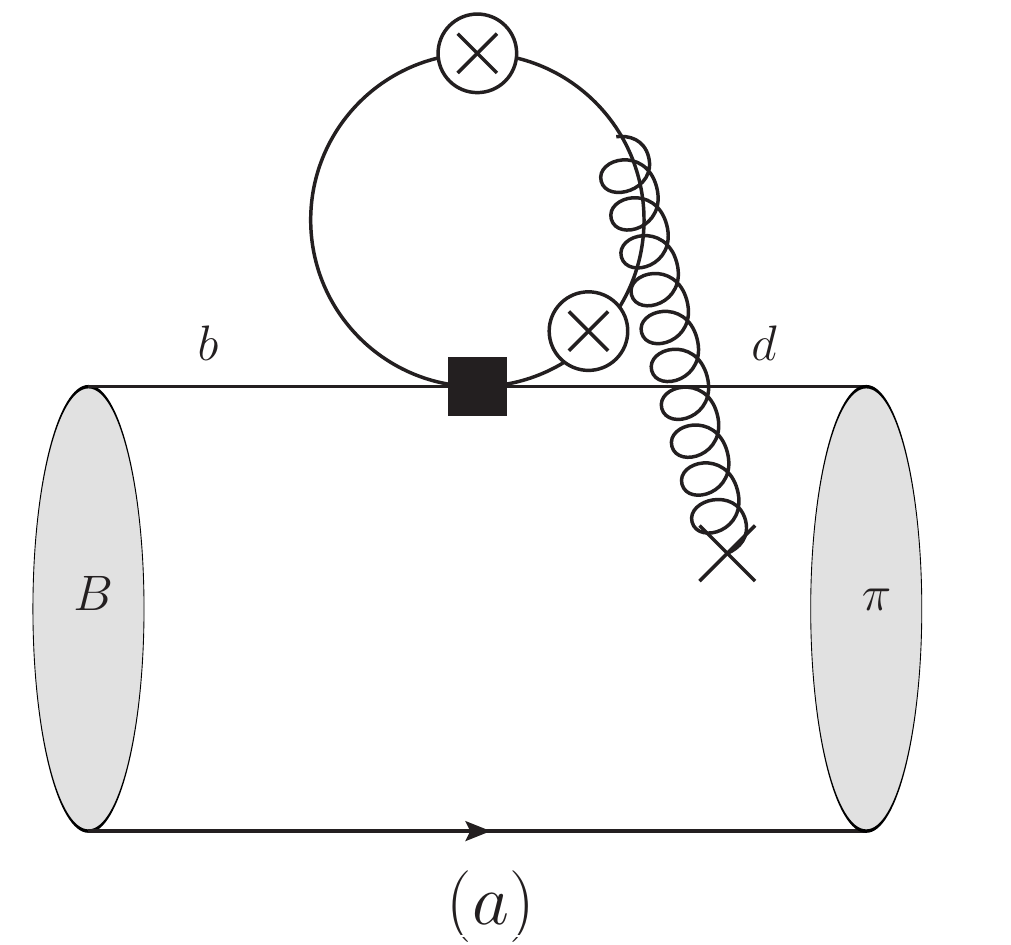}
\includegraphics[scale=0.42]{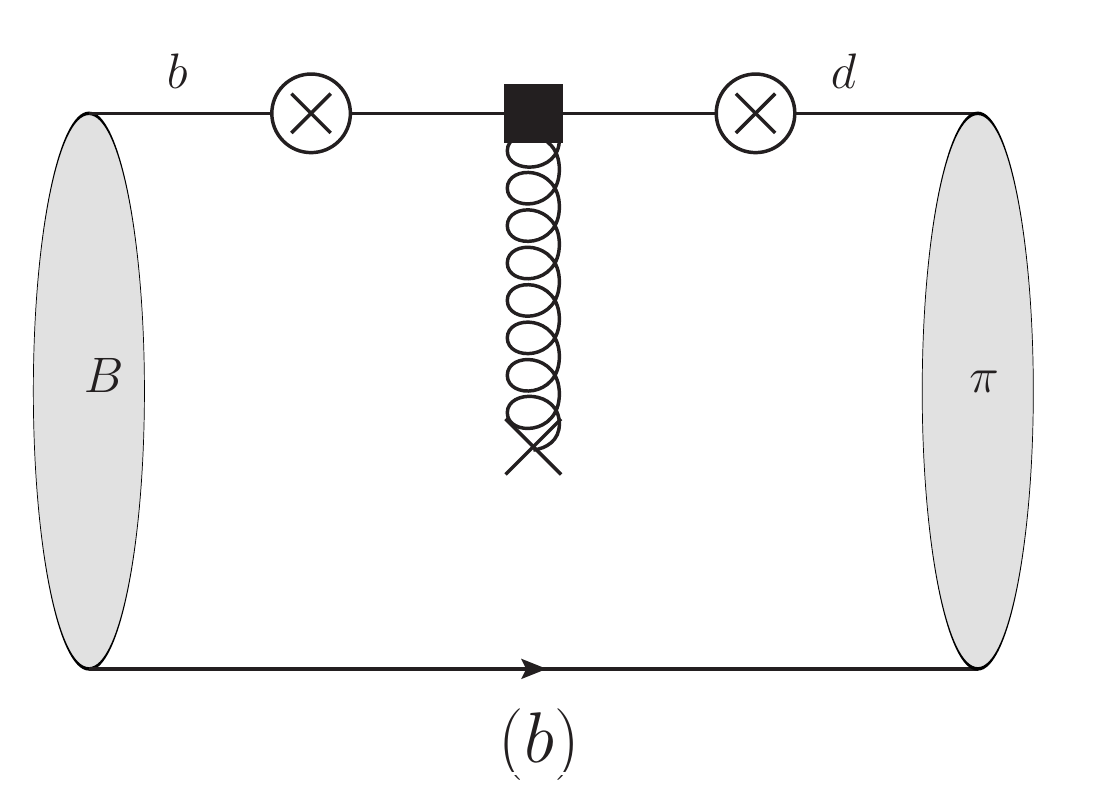}\\
\caption{ \it Nonfactorizable soft-gluon contributions to $B \to \pi \ell^+ \ell^-$
due to (a) the four-quark effective operators $O_{1,2}^{u,c}$ and $O_{3-6}$ 
and (b) the chromomagnetic $O_{8g}$  operator. The soft gluon
is represented by the gluon line with a cross. 
The rest of  notation is the same as in the previous figures.}
\label{fig:Nonfact-soft}
\end{figure}

We also take into account the nonfactorizable contributions to the 
amplitudes ${\cal H}^{(p)} (q^2)$ emerging due to a soft-gluon  
emission from the quark loops, as shown  in Fig.~\ref{fig:Nonfact-soft}. 
These hadronic matrix elements
cannot be reduced to a $B\to \pi$ form factor. It is also not possible 
to attribute the soft gluon to one of the hadrons in the $B\to \pi$ transition.
The soft-gluon contributions are nevertheless well defined 
at $q^2<0$  and $|q^2| \gg \Lambda_{QCD}^2$. As shown in Ref.~\cite{KMPW_charmloop},
their  suppression with respect to the factorizable loops is 
controlled by the powers of 
$1/(4m_c^2-q^2)$  and $1/|q^2|$, 
stemming, respectively, from the $c$-quark and massless loops with soft gluon. 
The corresponding hadronic matrix elements were  first calculated 
for the $c$-quark loops  in Ref.~\cite{KMPW_charmloop}  
for $B\to K^{(*)}\ell^+\ell^-$. The calculation was 
done  in two steps: 
(1) applying the light-cone OPE at deep spacelike $q^2$ for the 
quark loop with soft-gluon emission, and (2) calculating
the $B\to K^{(*)}$ hadronic matrix element of the emerging 
quark-antiquark-gluon operator using LCSRs with the $B$-meson 
three-particle DAs. Completing  this result to include the loops with 
all possible quark flavours  is straightforward and was already 
done for $B\to K\ell^+\ell^-$ in Ref.~\cite{KMW_BKll}; to obtain the 
corresponding contribution to ${\cal H}^{(c)} (q^2)$ in $B\to \pi\ell^+\ell^-$, we only need to replace the
kaon by the pion. The soft-gluon nonfactorizable contribution of 
the operator $O_1^{u}$ 
contributing to ${\cal H}^{(u)} (q^2)$, 
is also easily obtained. The result for this contribution 
is cast in a compact form:  
\begin{eqnarray}
{\cal H}^{(p)}_{\rm soft} (q^2) & = &  \frac 4 3 \left(\delta_{pc}C_1 + C_4 - 
C_6 \right)\tilde{A} (m_c^2, q^2) + 
\frac 2 3 \left( 2\delta_{pu}C_1+C_4 - C_6 \right) \tilde{A} (0, q^2) \nonumber \\
 & - & \frac 2 3 \left(C_3 + C_4 - C_6 \right) \left( \tilde{A} (m_s^2, q^2) + \tilde{A} (m_b^2, q^2) \right)\,. 
\end{eqnarray}
A cumbersome 
expression for the nonfactorizable hadronic matrix element 
$\tilde{A}(m_q^2, q^2)$ obtained from LCSR 
can be found  in Ref.~\cite{KMPW_charmloop},(see  Eq. (4.8) therein), where  
the dependence on the quark mass squared is explicitly shown
and is indicated in the above expression. 
To adjust this expression to the $B\to \pi \ell^+\ell^-$  transition,   
one has to replace the decay constant, meson mass  
and threshold parameter in this equation: $f_K\to f_\pi$, $m_K \to m_\pi$, 
$s_0^K\to s_0^\pi$, thus taking into account
the flavour $SU(3)$ violation.
In the sum rule for $\tilde{A}(m_q^2, q^2)$, we use the ansatz for 
the three-particle $B$-meson DAs
suggested in Ref.~\cite{KMO}, with the parameter 
$\lambda_E^2=3/2 \lambda_B^2$, directly related
to the inverse moment of the two-particle DA $\phi^+_B$ specified in 
Eq.~(\ref{eq:BDAansatz}).  

The soft-gluon contribution of the chromomagnetic operator $O_{8g}$  
is described by the diagram in Fig.~\ref{fig:Nonfact-soft}b,  where instead of 
the loop factor, one has a pointlike emission of the soft gluon field. One modifies
the correlation function  accordingly and arrives at the LCSR  
that was already derived in Ref.~\cite{KMW_BKll} and presented 
in Eq. (4.7) therein\footnote{We notice that in the related Eq. 
(4.4) a factor $C_{8g}$ on the r.h.s. is missing.}. Making the 
necessary replacements for  $B\to \pi\ell^+\ell^-$, we  obtain   
\begin{equation}
{\cal H}_{\rm soft, O_{8g}}^{(u)}(q^2) = {\cal H}_{\rm soft, O_{8g}}^{(c)}(q^2) =
\Big[{\cal H}^{(BK)}_{\rm soft, O_{8g}}(q^2)\Big]_{f_K\to f_\pi\,, m_K\to m_\pi,\,s_0^K \to s_0^\pi}\,.
\end{equation}
As in the case of $B\to K \ell^+\ell^-$ transitions, this contribution 
turns out to be very small.

\subsection{Nonfactorizable spectator scattering}

\begin{figure}\center
\includegraphics[scale=0.4]{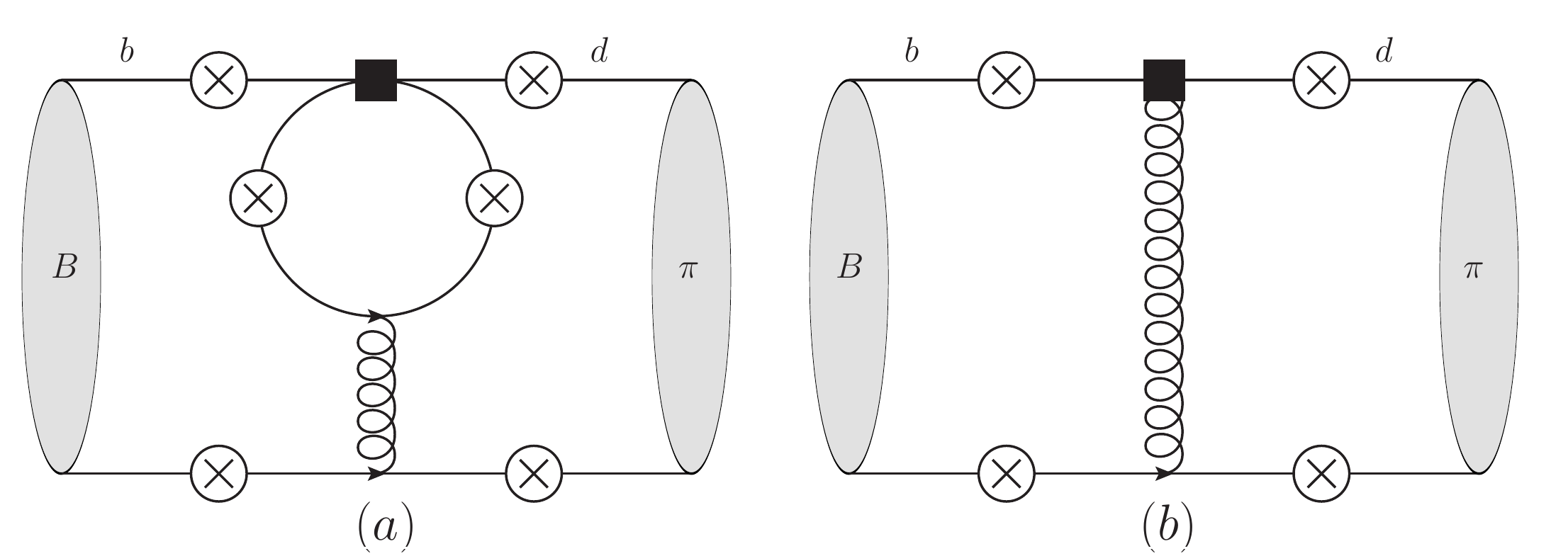}\\
\caption{\it Nonfactorizable spectator contributions to $B \to \pi \ell^+ \ell^-$
due to (a) the four-quark effective operators $O_{1,2}^{u,c}$, $O_{3-6}$ and (b)  
the chromomagnetic 
operator $O_{8g}$.}
\label{fig:Nonfact-spect}
\end{figure}

An important nonlocal contribution to the $B\to \pi \ell^+ \ell^-$ amplitude 
in NLO emerges due to a hard gluon emitted from 
the intermediate quark loop or from
the $O_{8g}$-operator vertex, and absorbed by the spectator quark in the $B \to \pi$
transition, as shown in Fig.~\ref{fig:Nonfact-spect}. 
Following \cite{KMW_BKll},  we will use the QCDF result \cite{BFS} 
for this contribution. The following 
expression is valid for both  $p=u$ and $p=c$ parts of the nonlocal amplitude:
\begin{eqnarray}
{\cal H}^{(p)}_{\rm nonf, spect}(q^2) & = & \frac{\alpha_s C_F}{32 \pi N_c} \frac{f_B f_\pi m_b}{m_B^2} \Biggl( \int\limits_0^\infty \frac{d\omega}{\omega} \phi_B^+ (\omega) \int\limits_0^1 d u \, \varphi_\pi (u) \, T_+^{(1),p} (u,\omega) + \nonumber \\
& + & \int\limits_0^\infty \frac{d\omega}{\omega} \phi_B^- (\omega) \int\limits_0^1 d u \, \varphi_\pi (u) \, T_-^{(1),p} (u,\omega) \Biggl)\,.
\label{eq:hardscat}
\end{eqnarray}
The hard-scattering kernels entering the above expression have the form:
\begin{eqnarray}
T_+^{(1),p} (u,\omega) & = & -\frac{m_B}{m_b} \Biggl(\frac 2 3 t_\parallel (u, m_c) (\delta_{pc}C_1 + C_4 - C_6)
- \frac 1 3 t_\parallel (u, m_b) (C_3 + C_4 - C_6) - \nonumber \\ 
& &  - \frac 1 3 t_\parallel (u, m_s) (C_3 + C_4 - C_6) + \frac 1 3 (2\delta_{pu}C_1+C_4 - C_6) t_\parallel (u,0) \Biggl), 
\end{eqnarray}
\begin{eqnarray}
T_-^{(1),p}(u,\omega) & = & - Q_q \frac{m_B \omega}{m_B \omega - q^2 - i \epsilon} \Biggl[ 
\frac{8 m_B}{3 m_b} \biggl(h(m_c^2, \bar u m_B^2 + u q^2) (\delta_{pc}C_1 + C_4 + C_6) + \nonumber  \\
& & + h(m_b^2, \bar u m_B^2 + u q^2) (C_3 + C_4 + C_6) + \nonumber \\
& & + h(0,\bar u m_B^2 + u q^2) (\delta_{pu}C_1 +C_3 + 3 C_4 + 3 C_6) -
\nonumber \\
& & - \frac 2 3 (C_3 - C_5 - 15 C_6) \biggr) + \frac{8 C_8^{\rm eff}}{\bar u + u q^2/m_B^2} \Biggr],
\end{eqnarray}
where
$Q_q$ is the electric charge of the spectator quark in the $B$-meson ($q = u,d$) and 
the functions $t_\parallel (u, m_q)$ and $h(m_q^2, q^2)$ can be found in Ref.~\cite{BFS}.

The two-particle $B$-meson DAs $\phi_B^\pm (\omega)$ are given in 
Eq.(\ref{eq:BDAansatz});
for the twist-2 pion DA we employ the standard Gegenbauer expansion:
\begin{eqnarray}
\varphi_\pi(u,\mu)= 6u(1-u)\Big(1+a_2^\pi (\mu) C_2^{(3/2)}(u) +a_4^\pi (\mu) C_4^{(3/2)}(u)\Big)\,.
\label{eq:varphi}
\end{eqnarray}
The fact that the amplitudes in (\ref{eq:hardscat}) depend on the charge of 
the spectator quark 
in  the $B$ meson  triggers another important contribution to the isospin
asymmetry in $B\to \pi \ell^+\ell^-$.

Summing up all contributions considered in this section, we obtain
the two nonlocal amplitudes in the adopted approximation:
\begin{eqnarray}
{\cal H}^{(p)}(q^2)&=&
{\cal H}^{(p)}_{\rm fact,LO}(q^2)+
{\cal H}^{(p)}_{\rm WA}(q^2)+
{\cal H}^{(p)}_{\rm fact, NLO}(q^2)
\nonumber\\
&+&{\cal H}^{(p)}_{\rm soft}(q^2)+
{\cal H}^{(p)}_{\rm soft, O_8}(q^2)+
{\cal H}^{(p)}_{\rm nonf, spect}(q^2)\,,\qquad  (p=u,c)\,.
\label{eq:Htot}
\end{eqnarray}

\section{Numerical analysis}
\label{sec:numerics} 
\begin{figure}[t]\center
\includegraphics[scale=1.0]{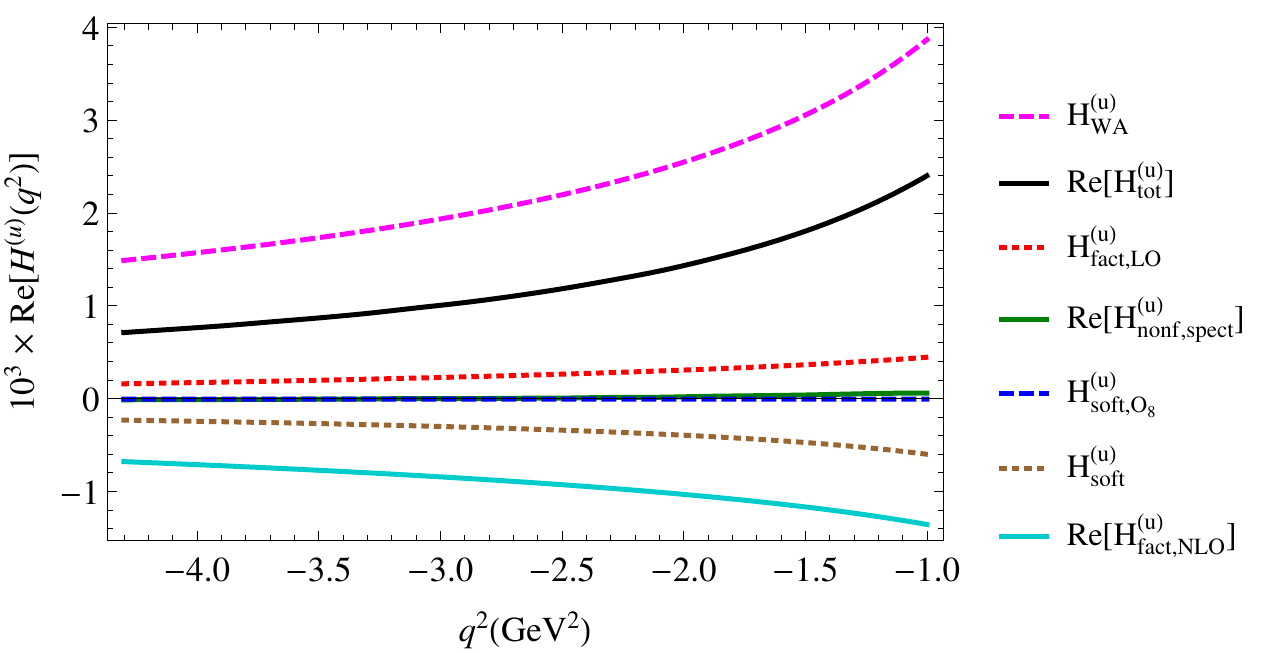} 
\includegraphics[scale=1.0]{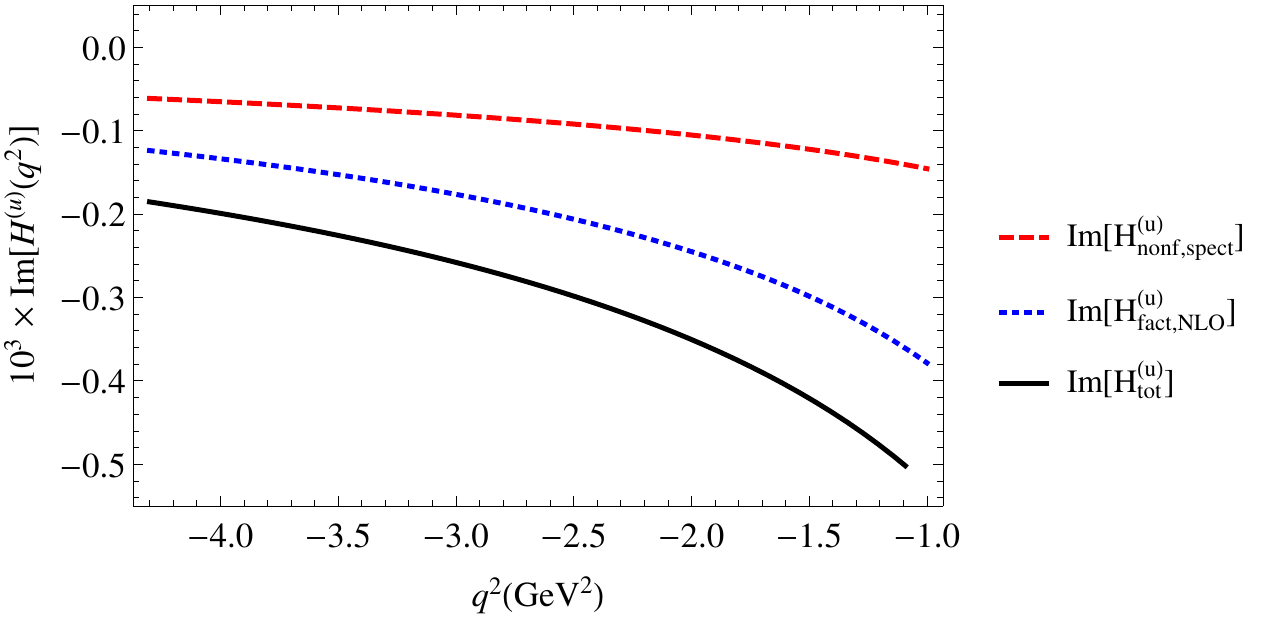}\\[2mm]
\caption{\it  Nonlocal amplitude ${\cal H}^{(u)}(q^2)$  
for $B^- \to \pi^- \ell^+ \ell^-$ at $q^2<0$
calculated at the central values of the input; 
the real (imaginary) part  is in the upper (lower) panel;
$H^{(u)}_{\rm tot} (q^2)$ is the sum of the separate contributions specified in Eq.(\ref{eq:Htot}). }
\label{fig:Hq2less0A1}
\end{figure}
\begin{figure}[t]\center
\includegraphics[scale=1.0]{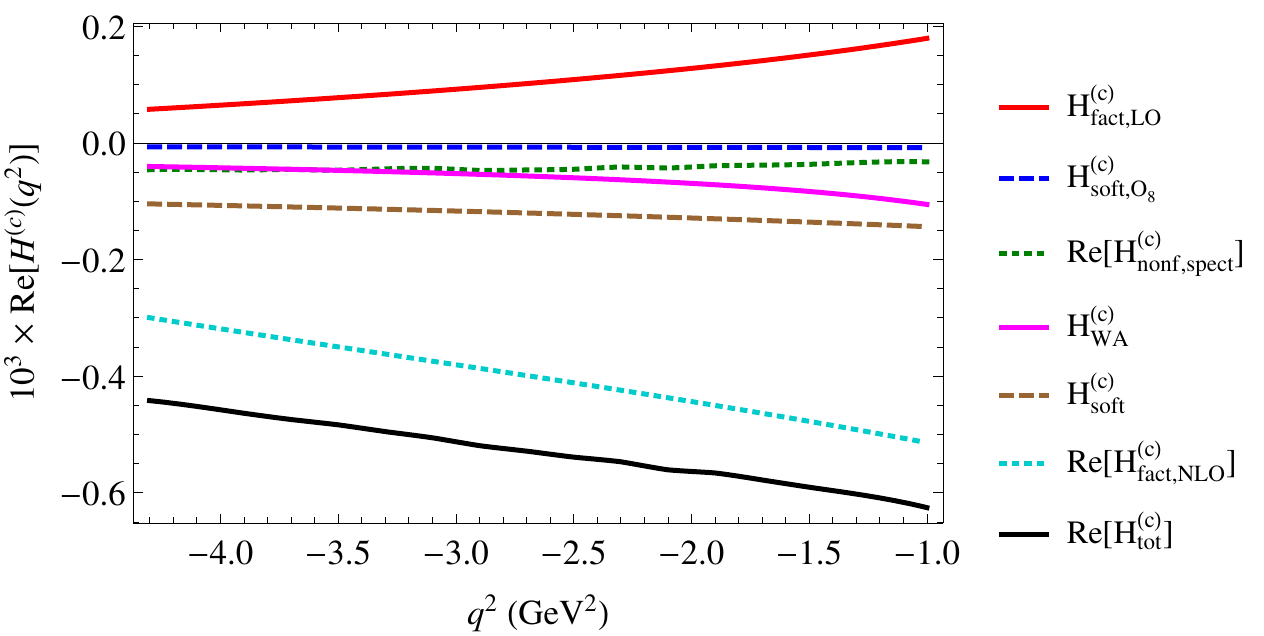}\\[2mm]
\includegraphics[scale=1.0]{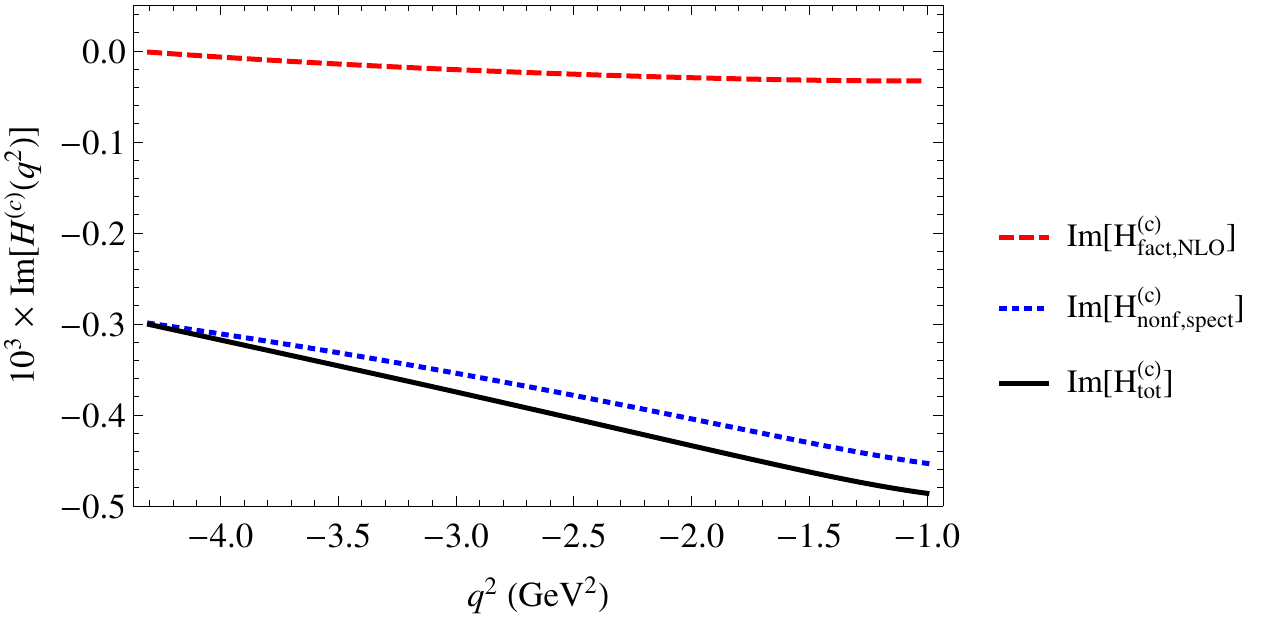} 
\caption{\it The same as in Fig.\ref{fig:Hq2less0A1} for the amplitude ${\cal H}^{(c)}(q^2)$  
for $B^- \to \pi^- \ell^+ \ell^-$. }
\label{fig:Hq2less0A2}
\end{figure}
Here we perform the numerical analysis
of the nonlocal amplitudes (\ref{eq:Htot}) at spacelike $q^2<0$,
more definitely, in the region  $1 \,\mbox{ GeV}^2\leq |q^2|
\lesssim 4\, \mbox{ GeV}^2 $,  
where the OPE  
and QCDF approximation  can be trusted. The input parameters and 
references to their source are listed 
in Table~\ref{tab:inputs}, 
the charged and neutral $B$-meson and pion 
masses are taken from \cite{PDG14}, and  
the numerical values of the Wilson coefficients 
are presented in the Appendix A.   
\begin{figure}[t]
\center
\includegraphics[scale=1.0]{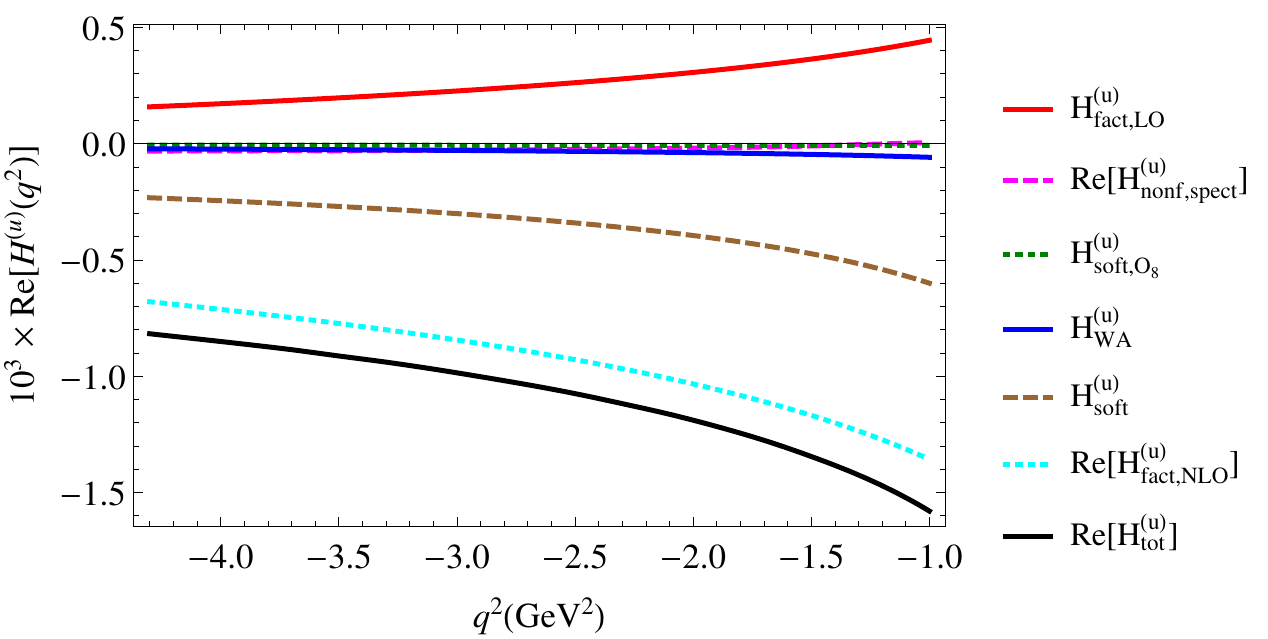} \\
\includegraphics[scale=1.0]{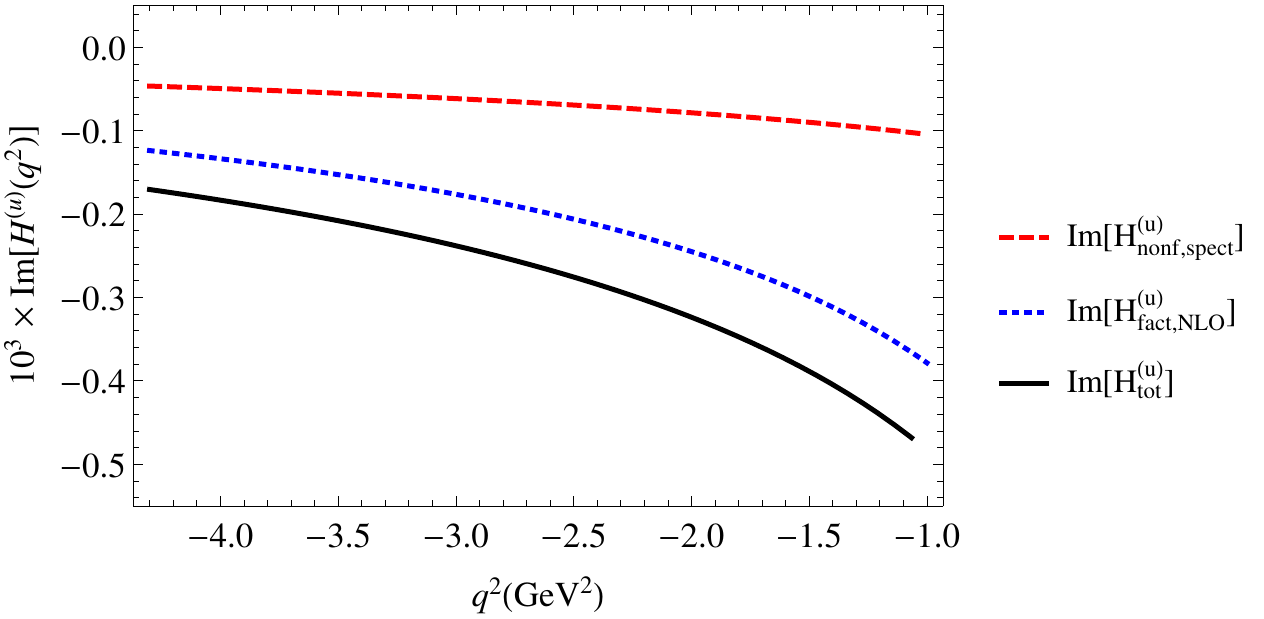}
\caption{\it The same as in Fig.\ref{fig:Hq2less0A1} for the amplitude ${\cal H}^{(u)}(q^2)$  
in $\bar{B}^0 \to \pi^0 \ell^+ \ell^-$.} 
\label{fig:Hq2less0B1}
\end{figure}
\begin{table}[h]
\begin{center}
\begin{tabular}{|l|l|}
\hline
Input parameter & [Ref.] \\
\hline
$\alpha_s (m_Z) = 0.1185 \pm 0.0006 $ & \\
$\overline{m}_c (\overline{m}_c) = (1.275 \pm 0.025)$ GeV & \\[-3mm]
$\overline{m}_b (\overline{m}_b)= (4.18 \pm 0.03)$ GeV & \Bigg\}\cite{PDG14}
\\[-4mm]
$m_s (2 {\rm GeV}) = (95 \pm 5)$ MeV\,~~ $\qquad$ & \\
\hline
$f_B = 207^{+17}_{-9}$ MeV & \cite{KPGR_2013} \\
$\lambda_B = (460 \pm 110) $ MeV & \cite{BIK} \\
\hline
$f_\pi = 130.4$ MeV & \cite{PDG14}\\
$a_2^\pi (1 {\rm GeV}) = 0.17 \pm 0.08 $
&\\
$a_4^\pi (1 {\rm GeV}) = 0.06 \pm 0.10 $& \cite{KMOW}\\
\hline
Sum rules in the pion channel &\\
$M^2 = 1.0\pm 0.5$ GeV$^2$\,,~ 
$s_0^{\pi} = 0.7 $ GeV$^2$ & \cite{KMO}\\
\hline
$f^+_{B\pi}(0)=0.307\pm 0.020$  &\\[-3mm]
$b_1^+= -1.31 \pm 0.42$ &  \Bigg\}\cite{IKMDvD}   \\[-4mm]
$b_2^+= -0.904 \pm 0.444$  &\\
\hline
\end{tabular}
\caption{\it Intervals of the input parameters used  in the calculation of ${\cal H}^{(u,c)}(q^2<0)$.}
\label{tab:inputs}
\end{center}
\end{table}
As a default renormalization and factorization 
scale we assume $\mu=3$ GeV, the same  as in Ref.~\cite{KMW_BKll}. It will be varied 
in the interval $2.5<\mu< 4.5 $~GeV to study the $\mu$ dependence.

For the vector $B\to \pi$ form factor 
the most recent update \cite{IKMDvD} of LCSR prediction is adopted, in a form 
fitted to the three-parameter BCL parameterization:
\begin{eqnarray}
    f^+_{B\pi}(q^2) = \frac{f^+_{B\pi}(0)}{1 - q^2/m_{B^*}^2}
\Bigg\{1 + b_1^+ \Bigg[z(q^2,t_0) - z(0,t_0)
- \frac{1}{3} \Big( z(q^2,t_0)^3 - z(0,t_0)^3\Big)\Bigg]
\nonumber \\
    + b_2^+ \Bigg[z(q^2,t_0)^2 - z(0,t_0)^2 + \frac{2}{3}\Big(z(q^2,t_0)^3
- z(0,t_0)^3\Big)\Bigg]\Bigg\}
\label{eq:fBpi}
\end{eqnarray}
with $t_0=(m_B+m_\pi)(\sqrt{m_B}-\sqrt{m_\pi})^2$,   where the normalization and shape parameters are presented in 
Table~\ref{tab:inputs}. The decay constant of $B$ meson is determined 
from two-point sum rules, where we use the recent analysis \cite{KPGR_2013};
the inverse moment of the $B$-meson DA is also represented
by the interval of QCD sum rule prediction \cite{BIK}. The intervals of Gegenbauer 
moments in the pion DAs used in the QCDF expressions are the same 
as in the LCSR for the $B\to \pi$ form factor \cite{IKMDvD,KMOW}. 
 
Substituting the central input in the expressions 
presented in the previous section, we calculate 
the two amplitudes ${\cal H}^{(u)}(q^2<0)$ and ${\cal H}^{(c)}(q^2<0)$
 for $B^-\to\pi^-\ell^+\ell^-$ and plot their real and imaginary parts 
in Figs.~\ref{fig:Hq2less0A1} 
and \ref{fig:Hq2less0A2}, respectively,
showing also the separate contributions.
For comparison, the amplitude ${\cal H}^{(u)}(q^2<0)$
is plotted in Fig.~\ref{fig:Hq2less0B1} for  $\bar{B}^0\to\pi^0\ell^+\ell^-$,
whereas the   amplitude ${\cal H}^{(c)}(q^2<0)$ for the latter mode
is numerically similar to the one for $B^-\to\pi^-\ell^+\ell^-$ and is not shown.

From the numerical analysis we can draw several conclusions:

(a) The contributions to ${\cal H}^{(c)} (q^2)$ 
are approximately  the same as the corresponding ones 
for the $B \to K \ell^+ \ell^-$ obtained in Ref.~\cite{KMW_BKll}, 
the differences reflect the violation of the flavour $SU(3)$ 
symmetry;  

(b) The contributions to ${\cal H}^{(u)} (q^2)$,
in $B^- \to \pi^- \ell^+ \ell^-$  presented 
in Fig. \ref{fig:Hq2less0A1} are clearly
dominated by the weak annihilation term, enhancing 
the real part of ${\cal H}^{(u)} (q^2)$ considerably.
This effect is less pronounced for  $\bar{B}^0 \to \pi^- \ell^+ \ell^-$,
as expected;

(c) The nonfactorizable soft-gluon contribution
due to the operators $O_{1,2}^{u,c}$ are important
and the corresponding contributions of $O_{3-6}$ are not negligible.
Meanwhile, the contribution due to the operator $O_{8g}$ 
with a soft-gluon is very small.

The uncertainties of the functions ${\cal H}^{(c)} (q^2<0)$ and ${\cal H}^{(u)} (q^2<0)$
are estimated  varying the inputs within their adopted intervals indicated in 
Table~\ref{tab:inputs}. 
The largest uncertainties originate from the variation of $f_B$, $\lambda_B$ 
and the correlated variation of the parameters of $f_{B\pi}^+$.
To stay on the conservative side, we neglect  possible 
correlations between the individual input entries in Table~\ref{tab:inputs}.
We also varied the renormalization/factorization scale  
around the default ``optimal'' value $\mu=3.0$ GeV. The results 
do not significantly change the estimated total uncertainty and we
therefore neglect the scale dependence in the error estimates performed below.

\section{Nonleptonic $B\to V \pi$ decay amplitudes}
\label{sec:nonlept}

We now turn to  weak nonleptonic $B \to V \pi$ decays
with neutral vector mesons 
$V = \rho^0, \omega, \phi, J/\psi, \psi (2S)$  in the final state.
The intervals of the absolute values of their amplitudes 
have to be estimated and used as  an 
input in the hadronic dispersion relations for ${\cal H}^{(u,c)}(q^2)$
to be fitted to the calculated ${\cal H}^{(u,c)}(q^2<0)$.

The amplitude of a  $B^- \to V \pi^-$  decay
is parametrized as:
\begin{eqnarray}
A(B^-\to V \pi^-)& =& \langle V(q)\pi^-(p)|H_{\rm eff(NL)}^{b\to d} | B^-(p+q)\rangle 
\nonumber\\
&=&\frac{4 G_F}{\sqrt 2} m_V(\varepsilon_V^* \cdot p) \bigg( \lambda_u A_{B^- V\pi^-}^u 
+\lambda_c A_{B^- V\pi^-}^c \bigg)\,,
\label{eq:BVpi}
\end{eqnarray}				
where $q$ ($\varepsilon_V$) is the 4-momentum
(polarization vector) of the vector meson with $q^2=m_V^2$.
For the charge-conjugated mode $B^+\to V\pi^+$ one has to replace
$\lambda_{u,c}\to \lambda_{u,c}^*$ 
in the above relation, whereas  the  hadronic amplitudes  
remain unchanged,  $A_{B^-V\pi^-}^{u,c}=A_{B^+V\pi^+}^{u,c}$. 
For the neutral $\bar{B}^0\to V\pi^0$ decay modes we denote the corresponding amplitudes 
as $A_{\bar{B}^0V\pi^0}^{u,c}=A_{B^0V\pi^0}^{u,c}$. 
The effective Hamiltonian $H_{\rm eff(NL)}^{b\to d}$ in Eq.~(\ref{eq:BVpi})
contains the operators $O_{1,2}^{u,c}$, $O_{3-6}$, $O_{8g}$ given in Appendix A, and 
we neglect the electroweak quark-penguin operators with  
$O(\alpha_{\rm em})$ suppressed Wilson coefficients.
From Eq.~(\ref{eq:BVpi}) one obtains the expression for the $CP$-averaged  width:
\begin{eqnarray}
\Gamma (B^\mp \to V \pi^\mp) 
\equiv \frac12 \big[\Gamma (B^- \to V \pi^-)+\Gamma (B^+\to V \pi^+)\big]
= \frac{G_F^2 \lambda^{3/2} (m_B^2, m_\pi^2, m_V^2)}{8 \pi m_B^3} 
\nonumber \\
\times\bigg(
|\lambda_u |^2|A_{B^-V\pi^-}^u|^2 +|\lambda_c|^2|A_{B^-V\pi^-}^c|^2 
+2|\lambda_u\lambda_c||A_{B^-V\pi^-}^u||A_{B^-V\pi^-}^c|\cos\delta \cos\Delta
\bigg)\,,
\label{eq:GamBVpi}
\end{eqnarray}
where $\lambda (a,b,c) = a^2 + b^2 + c^2 -2 a b - 2 a c - 2 b c$.
In the above,
we explicitly isolate the relative strong phase $\delta$ between 
the amplitudes  $A_{B^-V\pi^-}^u$ and $A_{B^-V\pi^-}^c$ and denote 
$\Delta  \equiv  {\rm arg}(\lambda_{u})- {\rm arg}(\lambda_{c})$.
The direct $CP$ asymmetry takes the following form:
\begin{eqnarray}
{\cal A}_{CP}(B^{\mp} \to V \pi^{\mp})
\equiv\frac{\Gamma (B^- \to V \pi^-)-\Gamma (B^+\to V \pi^+)}{2\Gamma(B^{\mp} \to V \pi^{\pm})}
\nonumber\\
= -2 \sin{\delta}\sin \Delta\bigg(
\left| \frac{\lambda_u A^u_{B^- V \pi^-}}{\lambda_c A^c_{B^- V \pi^-}}
\right|
+ \left |\frac{\lambda_c A^c_{B^- V \pi^-}}{\lambda_u A^u_{B^- V \pi^-}}\right|
+ 2 \cos \delta \cos \Delta \bigg)^{-1}\,.
\label{eq:Acp}
\end{eqnarray}
Analogous expressions are obtained for the neutral $B\to V \pi^0$  modes,
replacing $B^{-}\to \bar{B}^0$ and $B^{+}\to B^0$. 

For the input in the two dispersion relations for 
the amplitudes ${\cal H}^{(u)}(q^2)$
and   ${\cal H}^{(c)}(q^2)$ 
we need to separately determine 
the moduli of the hadronic amplitudes  $|A_{B^-V\pi^-}^{u}|$
and   $|A_{B^-V\pi^-}^{c}|$.  In principle, one can use the two 
observables in Eqs~(\ref{eq:BVpi})
and (\ref{eq:Acp}), but the presence of the third unknown
parameter, the relative strong phase, hinders the 
determination. Therefore, the situation is more complex here than for  
the nonleptonic $B\to V K$ decays 
used in the analysis of  $B\to K \ell^+\ell^-$ in Ref.~\cite{KMW_BKll}
 where  only the contributions proportional to $\lambda_c$
were retained  and the amplitudes $|A_{B V K}^{c}|$
were directly obtained from the measured $B\to V K$ branching fractions. 

On the other hand, there is a possibility to estimate
separate contributions 
to the nonleptonic amplitudes applying the QCDF approach
\cite{BBNS}. The latter is known to provide a reasonably good description of 
the charmless channels  $B^- \to \rho^0 \pi^- $ 
and $B^- \to \omega^0 \pi^- $.  Here we use the results
of Ref.~\cite{BN03} where the QCDF description for 
$B\to VP$ decays was elaborated  in detail.
The necessary expressions for the amplitude decomposition and the additional 
input parameters including the $B\to V $ form factors, 
decay constants and the Gegenbauer moments of the DAs of $V=\rho,\omega$ 
are collected in Appendix~B. The resulting absolute values of the 
amplitudes are presented in 
Table~\ref{Tab:ABVpi-u-c}. To check the validity of these estimates,
we calculated the observables (\ref{eq:BVpi})
and  (\ref{eq:Acp}), and compared the results
with the experiment and with the earlier predictions of Ref.~\cite{BN03} 
(see Table~\ref{Tab:BVpi-Br-exp-and-th}), observing a reasonable agreement. 
In the transition from the widths to branching fractions  
we use the lifetimes of $B$ mesons from Ref.~\cite{PDG14}.
\begin{table}[h]\center
\begin{tabular}{|l||c||c|}
\hline
Mode & $|A_{BV\pi}^u|$ & $|A_{BV\pi}^c|$  \\
\hline
$B^\mp\to \rho^0 \pi^\mp$ & $20.8^{+2.7}_{-2.3}$ & $1.3^{+1.1}_{-0.4}$ \\
\hline
$B^\mp\to \omega \pi^\mp$ & $19.1^{+2.7}_{-2.0} $ & $0.3^{+0.4}_{-0.1}$ \\
\hline
$B^\mp\to J/\psi \pi^\mp$ & $0.5^{+9.7}_{-0.5}$ & $29.2^{+1.4}_{-1.5} $  \\
\hline
$B^\mp\to \psi (2S)\pi^\mp$ & $3.5^{+6.7}_{-3.5} $ & $32.3^{+2.0}_{-2.1} $ \\
\hline
\end{tabular}
\begin{tabular}{|l||c||c|}
\hline
Mode & $|A_{BV\pi}^u|$ & $|A_{BV\pi}^c|$  \\
\hline
$B^0 \to \rho^0 \pi^0$ & $9.9^{+1.3}_{-1.4} $ & $0 $ \\
\hline
$B^0 \to \omega \pi^0$ & $0$ & $0$ \\
\hline
$B^0 \to J/\psi \pi^0$ & $0.3^{+6.9}_{-0.3}$ & $20.6^{+1.0}_{-1.1} $  \\
\hline
$B^0 \to \psi (2S)\pi^0$ & $2.4^{+4.7}_{-2.4} $ &  $22.8^{+1.4}_{-1.5} $ \\
\hline
\end{tabular}
\caption{ \it Inputs for the absolute values  
$|A_{BV\pi}^{u,c}|$ of the $B\to V\pi$ amplitudes (in MeV).}
\label{Tab:ABVpi-u-c}
\end{table}
Note that in the dispersion relations 
we will not isolate the intermediate $\phi$-meson pole, 
hence we do not specify  the 
$B\to \phi \pi$ nonleptonic amplitudes here.
These decays originate either due to the $q=s$ part 
of  the quark-penguin operators $O_{3-6}$ 
with suppressed Wilson coefficients, or due to the $O_{1,2}$ 
operators combined with a transition 
via intermediate  gluons into $\bar{s}s$ state. The latter is 
OZI suppressed (cf. the smallness of $\phi$ decays into
pions). The measured upper limit ${\rm BR}(B^{-}\to \phi \pi^{-})<1.5\times 10^{-7}$  
\cite{PDG14}, being significantly smaller than the measured branching fractions of 
$B^{-}\to \rho (\omega) \pi^{-}$ decays (see Table~\ref{Tab:BVpi-Br-exp-and-th}) 
convinces us that the intermediate $\phi$-meson  contribution 
to $B\to \pi \ell^+ \ell^-$ 
is small. 
Furthermore  we do not separate 
the radial excitations $\rho',...,\omega',...$, 
approximating their contributions 
to the  hadronic spectral density by the  
quark-hadron duality ansatz; hence we do not need to consider 
here the nonleptonic decays  of the type $B\to \rho'(1450)\pi$. 

\begin{table}[h]\center
\begin{tabular}{|l|c|c|c|c|}
\hline
Channel& Observable &Experiment & QCDF \cite{BN03}& QCDF, this work  \\
\hline
$B^\pm \to \rho^0 \pi^\pm$& ${\rm BR}\times 10^{6}$& $ 8.3 \pm 1.2  $ & $11.9^{+7.8}_{-6.1}
$ &
$ 9.5 ^{+2.9}_{-1.8}$ \\
& ${\cal A}_{CP}$ & $0.18^{+0.09}_{-0.17} $& $\phantom{-}0.04 \pm 0.19$ & $ 0.09 \pm 0.17$
\\
\hline
$B^\pm \to \omega^0 \pi^\pm$&${\rm BR}\times 10^{6}$ & $6.9 \pm 0.5 $ & $8.8^{+5.4}_{-4.3}
$&
$ 8.9^{+2.6}_{-1.6}$ \\
& ${\cal A}_{CP}$ & $-0.04 \pm 0.06 $ & $-0.02 \pm 0.04$ & $ -0.06 \pm 0.06 $\\
\hline
$B^0 \to \rho^0 \pi^0$& ${\rm BR}\times 10^{6}$& $ 2.0 \pm 0.5  $ & $0.4^{+1.1}_{-0.4} $ &
$ 0.2 ^{+0.4}_{-0.1}$ \\
& ${\cal A}_{CP}$ & --- & $-0.16^{+0.26}_{-0.32}$ & $ 0.24^{+0.36}_{-0.31}$ \\
\hline
$B^0 \to \omega^0 \pi^0$ & ${\rm BR}\times 10^{6}$ & $ < 0.5$ & $0.01^{+0.04}_{-0.01} $&
$ 0.01^{+0.06}_{-0.01}$ \\
& ${\cal A}_{CP}$ & --- & --- & $ -0.94^{+0.87}_{-0.04} $\\
\hline
\end{tabular}
\caption{\it Comparison of the experimental data \cite{PDG14} 
and theoretical predictions for the observables  in 
the nonleptonic $B \to \rho(\omega) \pi$ decays.}
\label{Tab:BVpi-Br-exp-and-th}
\end{table}
For the neutral $B^0 \to V \pi^0$  modes,
 the QCDF prediction \cite{BN03} fails to predict the partial width 
$B^0 \to \rho^0 \pi^0$, the experimental value 
being significantly larger. Without going into more detailed 
discussion of this problem, guided by the hierarchy of amplitudes
in the charged mode, $A_{B^0\rho^0\pi^0}^c\ll A_{B^0\rho^0\pi^0}^u$,
we simply assume  $A_{B^0\rho^0\pi^0}^c = 0$ and 
extract  $|A_{B^0\rho^0\pi^0}^u|$ from the measured 
partial width employing  Eq.~(\ref{eq:GamBVpi}). 
For the $B^{0}\to \omega \pi^{0}$ mode
only the upper limit on the branching fraction is available  
\cite{PDG14}, indicating that this decay amplitude is suppressed
in comparison to the other modes, hence
we put $A_{B^0 \omega \pi^0}^u \approx A_{B^0 \omega \pi^0}^c \approx 0$
as it specified in Table \ref{Tab:ABVpi-u-c}.

Turning to the charmonium channels $B \to \psi  \pi$, where $\psi = J/\psi, \psi(2S)$, we do not expect
the QCDF approach to work there due to a heavy final state 
and enhanced nonfactorizable, power suppressed effects (see, e.g., the discussion
in Ref.~\cite{KMPW_charmloop}). On the other hand, one
anticipates  that  these nonleptonic decays are dominated by 
the emission topology   due to the operators $O_{1,2}^c$ 
with large Wilson coefficients 
and  a small admixture of $O_{3-6}$ ($q=c$).
The contributions of the operators $O_{1,2}^{u}$ and $O_{3-6}$ $(q\neq c)$ 
are expected to be strongly suppressed. Theoretical 
estimates for the analogous  contributions to  $B\to \psi K$ transitions  
(see, e.g., \cite{Mannel_etal} and \cite{Nierste_etal15}) yield 
the amplitudes of  gluonic transitions of light-quark loops to charmonium 
states  at the level of $10^{-3}$ of the dominant contributions 
of $O_{1,2}^{c}$ operators.  With an extra Cabibbo enhancement of 
the $\lambda_u$ terms in $B\to \psi \pi$ with respect to $B\to \psi K$, 
still a considerable suppression remains.
Hence we expect that $|A_{B^-\psi\pi^-}^u|\ll |A_{B^-\psi\pi^-}^c|$. 
In this situation
the relative strong phase does not considerably
influence the  extraction of the large $\sim \lambda_c$ term  ,
whereas the uncertainty of the small  $\sim \lambda_u$ term
is tolerable.  
Therefore, we use the current experimental data on the 
branching fractions and $CP$-asymmetries
of the above decays \cite{PDG14} 
and perform the fit of these data to the Eqs. (\ref{eq:GamBVpi})
and (\ref{eq:Acp}), extracting the absolute values of the amplitudes
$|A_{B^-\psi\pi^-}^u|$ and $|A_{B^-\psi\pi^-}^u|$ 
and allowing the relative phase $\delta$
to change from 0 to $2 \pi$. The resulting intervals are presented in Table~ 
\ref{Tab:ABVpi-u-c}. Finally, for the neutral $\bar{B}^0\to \psi \pi^0 $  modes,
we make use of the isospin symmetry relation:
$A_{B^0 \psi \pi^0}^{u,c} 
\simeq 1/\sqrt{2}A_{B^- \psi \pi^-}^{u,c} $,
since for the dominant  $\psi$ emission mechanism of these decays 
there is only one independent isospin amplitude. 
This assumption is supported by the measurement \cite{PDG14} yielding  
$\Gamma(B^0 \to J/\psi \pi^0) \simeq 1/2 \Gamma(B^+ \to J/\psi \pi^+)$ .
The resulting estimates are also presented in Table~\ref{Tab:ABVpi-u-c}.

\section{Hadronic dispersion relations}
\label{sec:disp}
 Following 
Refs.\cite{KMPW_charmloop,KMW_BKll}, 
the invariant amplitudes 
${\cal H}^{(u)} (q^2)$ and ${\cal H}^{(c)} (q^2)$ 
are represented in a form of 
hadronic dispersion relations in the variable $q^2$, 
inserting the total set of hadronic intermediate 
states between the electromagnetic current and the effective operators
in the correlation functions (\ref{eq:corr}):
\begin{eqnarray}
\label{eq:disp1}
{\cal H}^{(p)} (q^2)- {\cal H}^{(p)} (q_0^2) & = & 
(q^2 -q^2_0)\Biggr[\sum\limits_{V=\rho,\omega,J/\psi,\psi(2S)} 
\frac{k_V f_V A_{BV\pi}^{p}}{(m_V^2 - q_0^2 )(m_V^2 -q^2 - i m_V \Gamma_V^{\rm tot})}
\nonumber \\
& & + \int\limits_{s_h}^\infty d s \frac{\rho_h^{(p)}(s)}{(s-q_0^2)(s-q^2 - i \epsilon)}\Biggl] 
\,,\quad (p=u,c)\,,
\end{eqnarray}
where the ground-state vector mesons (except $\phi$) are isolated 
and the integral describes  the contribution of  excited 
and continuum contributions starting from 
$s_h=4m_\pi^2$, the lowest hadronic threshold
\footnote{ Note however that a part of the 
2-pion continuum contribution in this region  is effectively absorbed in 
the $\rho$ meson total width (for more details see, e.g., \cite{BruchKhK}).}.
To achieve a better convergence, we implement one subtraction 
at $q_0^2=-1.0 $ GeV$^2$ in Eq.~(\ref{eq:disp1}).
In the above, the masses and total widths of the vector mesons $V=\rho^0,\omega,
J/\psi,\psi(2S)$ are taken 
from Ref.~\cite{PDG14}. Their decay constants are defined as  
\begin{equation}
\label{ME:0-V}
\langle 0 | j^{{\rm em}, \, \mu} | V (q) \rangle = k_V m_V \varepsilon_{V}^\mu (q) f_V\,,
\end{equation}
where the coefficients $k_V$  are determined  by the valence-quark content of $V$ and the quark charges: $k_\rho = 1 /\sqrt{2}$, $k_\omega = 1/(3 \sqrt 2)$ and $k_{J/\psi} = k_{\psi (2S)} = 2/ 3$. The numerical values of $f_V$ are fixed from
the measured \cite{PDG14} leptonic widths $\Gamma(V\to \ell^+\ell^-)$ yielding 
$f_\rho = 221 $ MeV,  $f_\omega = 195 $ MeV, 
$f_{J/\psi} = 416 $ MeV and $f_{\psi(2S)} = 297 $ MeV. 
The absolute values of the amplitudes $A_{BV\pi}^{u}$  and  
$A_{BV\pi}^{c}$ obtained from the analysis of  nonleptonic decays 
in the previous section are taken from Table~\ref{Tab:ABVpi-u-c}. 

At $q^2<0$, more specifically, in the region  
$-4.0 ~\mbox{GeV}^2 \leq q^2\leq -1.0 ~\mbox{GeV}^2$, 
we substitute in the l.h.s of the relations (\ref{eq:disp1})
the result for $H^{(p)} (q^2)$ specified in Eq.~(\ref{eq:Htot}),
calculating simultaneously the subtraction terms at $q_0^2=-1.0~\mbox{GeV}^2$.  
The task is then to fit the free parameters on the r.h.s. of the 
hadronic dispersion relations. 
Importantly, each $V$-pole residue in  Eq.~(\ref{eq:disp1}) for $p=u$ or $p=c$ 
has a relative phase with respect to the other vector-meson
contributions and to the integral over  $\rho^{(p)}_h(s)$.
These phases should match 
the imaginary part of the  calculated l.h.s. of the dispersion relation.  
As explained in Ref.~\cite{KMW_BKll}, the phases  
emerge due to the intermediate on-shell hadronic states in the variable
$(p+q)^2=m_B^2$  and are not related to the analytical continuation in the variable $q^2$.

Note that  the relative strong phase between 
the amplitudes $A_{BV\pi}^{u}$  and $A_{BV\pi}^{c}$ contributing to the $B\to V \pi$  
nonleptonic amplitude, although of the same origin, is a different  
quantity,  because in each of dispersion relations (\ref{eq:disp1}) only 
one of these amplitudes enter. On the other hand, calculating 
the phases of nonleptonic amplitudes within a theoretical framework,
such as QCDF, it is possible to estimate the relative phase between, 
say, $A_{B\rho\pi}^{u}$ and $A_{B\omega\pi}^{u}$.

In what follows, we attribute a phase 
to each $V$ -pole term:
\begin{equation}
A_{BV\pi}^{p}=|A_{BV\pi}^{p}|\exp(i\delta_{BV\pi}^{(p)})\,.
\label{eq:phase}
\end{equation}
To reduce the number of free parameters, 
we fix the phase differences: 
\begin{equation}
\delta_{B^-\rho^0\pi^-}^{(u)}-\delta_{B^-\omega\pi^-}^{(u)}=0.033, ~~
\delta_{B^-\rho^0\pi^-}^{(c)}-\delta_{B^-\omega\pi^-}^{(c)}=-3.65,
\label{phasediff}
\end{equation}
calculating it from QCDF, as explained in the previous section. Note that for the 
neutral mode the contribution of the $\omega$-meson is neglected
and the corresponding difference is irrelevant.
The three remaining phases $\delta_{\rho}^{(p)}$, $\delta_{J/\psi}^{(p)}$ and 
 $\delta_{\psi(2S)}^{(p)}$ for each $p=u,c$ are included into the set of fit parameters.
This set will be completed below by the fit parameters of the integrals 
over $\rho_h^{(p)}(s)$. 
Furthermore, we adopt the Breit-Wigner 
form  of the vector meson contributions in (\ref{eq:disp1})   with 
an energy-dependent total width for the broad $\rho$-resonance 
so that it vanishes at $q^2<4m_\pi^2$ and adopting constant 
total widths  $\Gamma_V^{\rm tot}$ for the remaining narrow resonances.

To complete the ansatz for the  
hadronic dispersion relations, we have to specify  
the integrals over the hadronic spectral densities of excited and continuum states
$\rho^{(u,c)} (s)$ in Eq.(\ref{eq:disp1}).
In the region below the open charm threshold, $q^2=s\leq 4m_D^2$,
apart from the two narrow charmonium resonances $J/\psi$ and $\psi(2S)$, 
only the intermediate states
with light quark-antiquark flavour content and spin-parity $1^-$ 
contribute.  We make extensive use of the standard
quark-hadron duality ansatz employed in 
the  QCD sum rules \cite{SVZ} 
for the vector-meson channels. The integral over 
the hadronic spectral density $\rho_h^{(p)}(s)$  including 
$\rho',\omega',...$ and continuum
states with the $\rho$ and $\omega$ quantum numbers is replaced
by the spectral density calculated from OPE:
\begin{eqnarray}
\rho^{(p)}_h (s)\theta(s-s_h) &\simeq&\frac 1 \pi \left( {\rm Im} {\cal H}_{\rm fact,LO \{u,d \}}^{(p)} (s)  + {\rm Im} {\cal H}_{\rm WA}^{(p)} (s) \right) \theta(s - s_0)
\nonumber
\\
&+&\frac{1}{\pi} {\rm Im} {\cal H}_{\rm fact,LO\{s\}}^{(p)} (s) 
\theta (s -\tilde{s_0})   \,, ~~~(p=u,c;~ s<4m_D^2)\,,
\label{eq:rhou}
\end{eqnarray}
where only the LO contributions are taken into account, including the leading-order 
quark loops and weak annihilation diagrams. The indices $\{u,d\}$ and $\{s\}$ 
mean that only the diagrams with $u,d$ and $s$ quarks, respectively, are taken
into account.  
The duality threshold $s_0\simeq 1.5$ GeV$^2$ is chosen in accordance 
with the analysis of QCD sum rules in the light vector-meson channels. 
In the contribution of the intermediate $\bar{s}s$ hadronic states 
to the spectral density $\rho^{(p)}_h (s)$ (the last term in (\ref{eq:rhou}))   
we include also the small $\phi$-meson pole contribution. This is reflected by the 
choice of a lower effective threshold parameter  $\tilde s_0 =4m_K^2\simeq 1.0$ GeV$^2$. Taking at $s=(q^2+i\epsilon)$ the imaginary parts of the loop function
(\ref{eq:mloop}):
\begin{equation}
\label{LO-Im}
\frac{1}{\pi}{\rm Im}\,g(q^2+i\epsilon,m_q^2)=
\left(1+\frac{2m_q^2}{q^2} \right) \sqrt{1-\frac{4m_q^2}{q^2}}\,\theta (q^2 -4m_q^2)
\end{equation}
and of the WA contribution (\ref{WA-neg-qsq}):
\begin{equation}
\label{WA-Im}
\frac{1}{\pi}{\rm Im} {\cal H}_{\rm WA}^{(p)} (q^2+ i \epsilon ) = 
 \frac{Q_qf_B f_\pi}{2 N_c \lambda_B m_B}
e^{-q^2/(\lambda_B m_B)} \tilde C_{\rm WA}^{p} \theta (s)\,,
\end{equation}
we obtain   
\begin{eqnarray}
\rho^{(u)}_{LO} (s) & = & 
\theta(s - s_0) \biggl[ \frac{1}{24 \pi^2} \left(\frac 2 3 C_1 + 2 C_2 + C_3 + 
\frac{C_4} 3 + C_5 + \frac{C_6} 3 \right) f^+_{B\pi} (s) + 
\nonumber \\
& + & Q_q \frac{f_B f_\pi}{2 N_c \lambda_B m_B} e^{-s/(\lambda_B m_B)}
\left(\delta_{qu} (C_2 + 3 C_1) + \delta_{qd} (C_1 + 3 C_2) 
+ C_3 + 3 C_4 \right) \biggl]  
\nonumber \\  
& - & 
\frac{1}{24 \pi^3} \left(\frac 4 3 C_3 + \frac 4 3 C_4 + C_5 + \frac{C_6}{3} \right)
{\rm Im}\,g(s, m_s^2) f^+_{B\pi} (s)\theta (s - \tilde s_0) 
\label{eq:rhou1}
\end{eqnarray}
and a similar expression: 
\begin{eqnarray}
\rho^{(c)}_{LO} (s) 
& = & \biggl[ \frac{1}{24 \pi^2} \left(C_3 + 
\frac{C_4} 3 + C_5 + \frac{C_6} 3 \right) f^+_{B\pi} (s) + 
\nonumber \\
& + & Q_q \frac{f_B f_\pi}{2 N_c \lambda_B m_B}
e^{-s/(\lambda_B m_B)} \left( 
C_3 + 3 C_4 \right) \biggl] \theta(s - s_0) 
\nonumber \\  
& - & 
\frac{1}{24 \pi^3} \left(\frac 4 3 C_3 + \frac 4 3 C_4 + C_5 + \frac{C_6}{3} \right)
 {\rm Im}\,g(s, m_s^2) f^+_{B\pi} (s)\theta (s - \tilde s_0).
\label{eq:rhoc}
\end{eqnarray}   
The two above expressions specify the adopted ansatz 
(\ref{eq:rhou}) at
$s_h<s<4m_D^2$. Note that in the LO
approximation, the spectral densities $\rho^{(p)}_h (s)$ are real functions. 
Following Ref.~\cite{KMW_BKll} we slightly modify the denominator in the dispersion integral
over $s_0<s<4m_D^2$  
replacing $s-q^2-i\epsilon \to s-q^2-i\sqrt{s}\Gamma_{\rm eff}(s,q^2)$
where $\Gamma_{\rm eff} (s,q^2) = \gamma \sqrt{s} \Theta(q^2 - 4 m_\pi^2), \, \gamma = 0.2$
is the effective energy-dependent width, where the $\theta$ function ensures that 
this width is absent at negative $q^2$. This modification 
allows one to transform the smooth duality-driven spectral density towards 
more realistic series of equidistant vector mesons (cf. the model for the pion
timelike form factor used in Ref.~\cite{BruchKhK}).
The addition of NLO corrections to the LO approximation for the 
duality ansatz remains a difficult task for a future improvement, 
involving  a calculation of the spectral densities of the diagrams in 
Figs.~\ref{fig:NLO-fact} and \ref{fig:Nonfact-spect}. 

The spectral densities $\rho_h^{(p)} (s)$  
above the open charm threshold, $s>4m_D^2$, 
contain a complicated 
overlap of broad charmonium resonances and open-charm states, together 
with the light-quark contributions. Moreover, starting from $s=(m_B+m_\pi)^2$ 
the on-shell intermediate states with $b$ flavour related to the
imaginary part of the $B\to \pi$ form factor in the timelike region 
also contribute. Hence, a duality-based parameterization of the $s>4m_D^2$ 
part of the integral over  $\rho_h^{(p)} (s)$  
will not adequately reflect the complicated resonance-continuum
structure of the hadronic spectral density. 
On  the other hand, we only 
need this part of the integral at relatively small $q^2<m_{J/\psi}^2$, hence 
following  Ref.~\cite{KMW_BKll},  we use a simple expansion in the powers 
of $q^2/4 m_D^2$, truncating it at the first order:
\begin{equation}
\int\limits_{4 m_D^2}^\infty d s \frac{\rho^{(p)}(s)}{(s-q_0^2)(s-q^2 - i \epsilon)}
\simeq a_p + b_p \frac{q^2}{4 m_D^2}, \quad p = u, c,
\label{eq:dispintlargeq2}
\end{equation}
where $a_{u,c}=|a_{u,c}|e^{i\phi_a}$ and $b_{u,c}=|b_{u,c}|e^{i\phi_b}$ 
are two unknown complex parameters.
Note that in Ref.~\cite{KMW_BKll} other parameterizations of the 
dispersion integral were also probed, and the results in the large recoil 
region were numerically close to the ones obtained with Eq.~(\ref{eq:dispintlargeq2}),
hence we will only consider this choice. 

\begin{figure}\center
\includegraphics[scale=0.8]{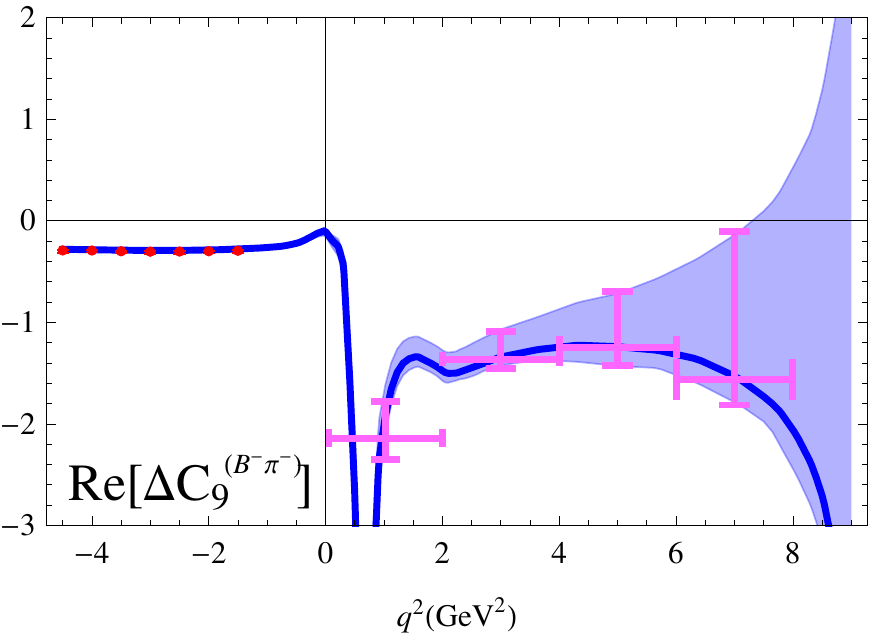}
\includegraphics[scale=0.8]{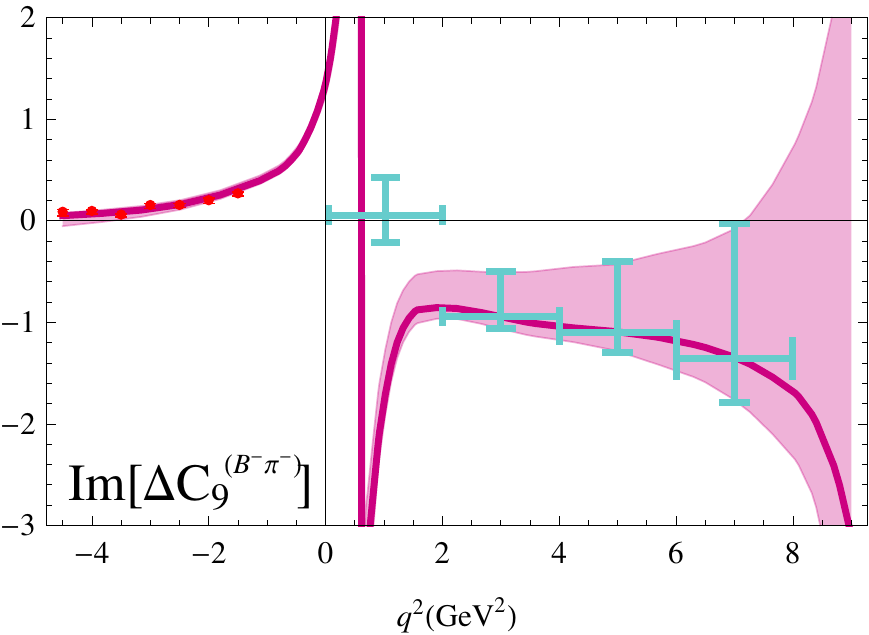}\\[2mm]
\includegraphics[scale=0.8]{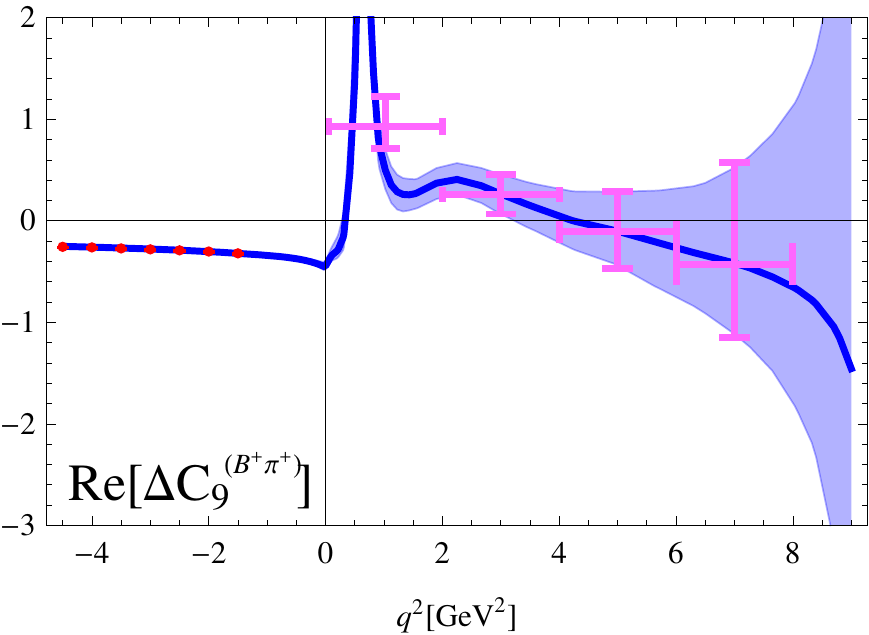}
\includegraphics[scale=0.8]{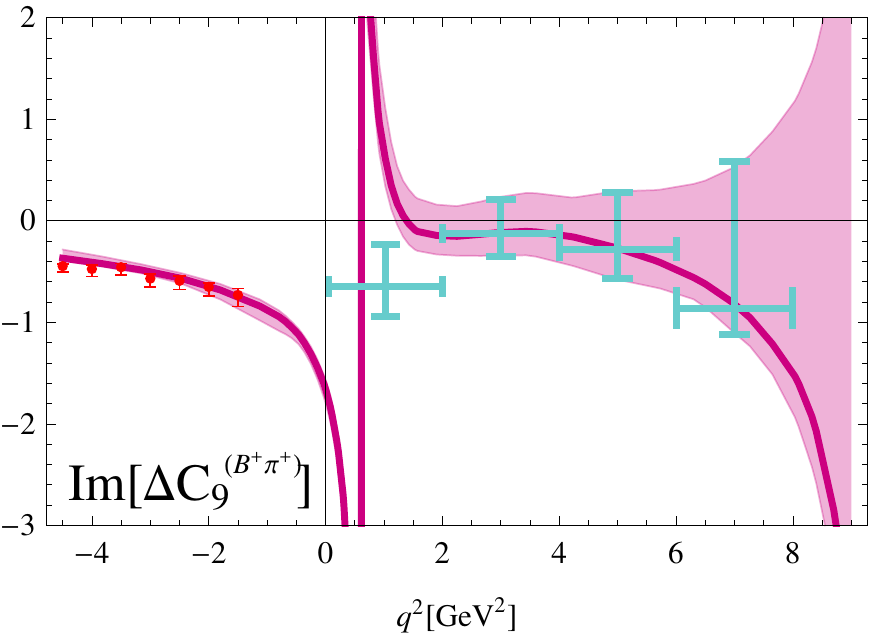}
\caption{\it The real part (upper left, blue online) and imaginary part 
(upper right, red online) of $\Delta C^{B\pi}_9 (q^2)$  
for $B^- \to \pi^-\ell^+ \ell^-$. 
The solid line is  the dispersion relation fitted to  the results 
calculated at $q^2<0$ at the central input. The shaded area indicates 
the estimated 68\% C.L. uncertainties obtained in the fit to the ``data" points
(dots,red online) at $q^2<0$. The values of $\Delta C^{B\pi}_9 (q^2)$  
averaged over the $q^2$-bins are also shown. The lower panel 
contains the same plots for $B^+ \to \pi^+\ell^+ \ell^-$. }
\label{fig:DC9charged}
\end{figure}

\begin{figure}\center
\includegraphics[scale=0.8]{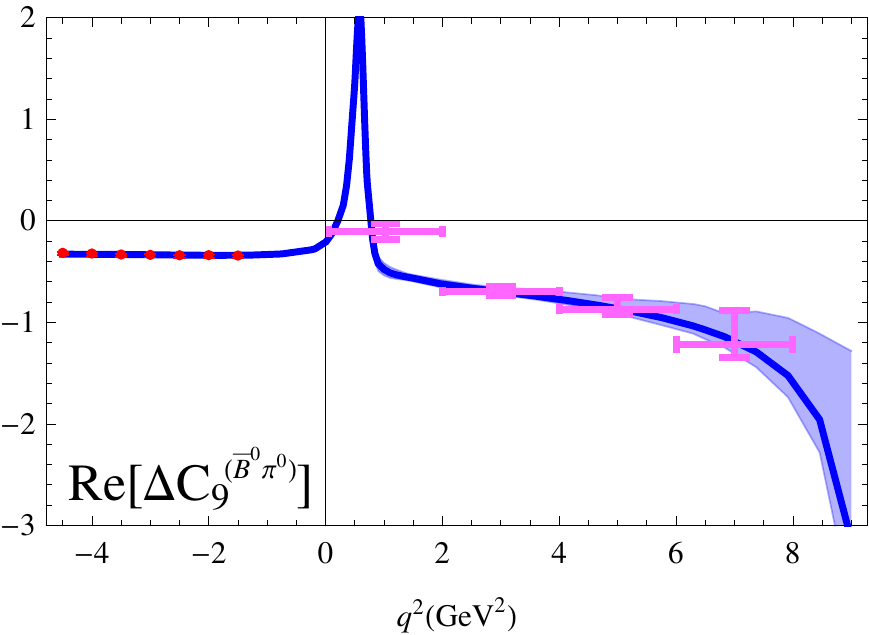}
\includegraphics[scale=0.8]{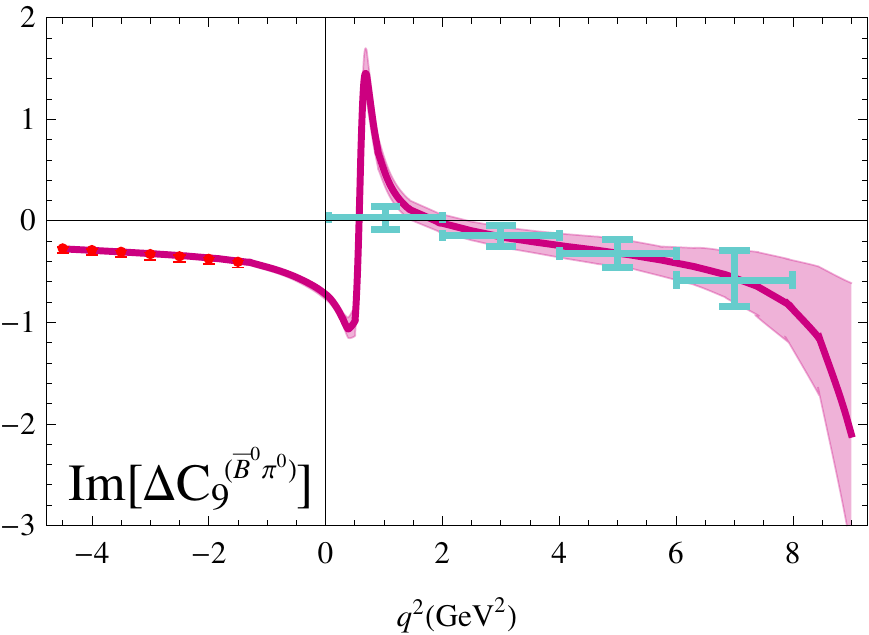}\\[2mm]
\includegraphics[scale=0.8]{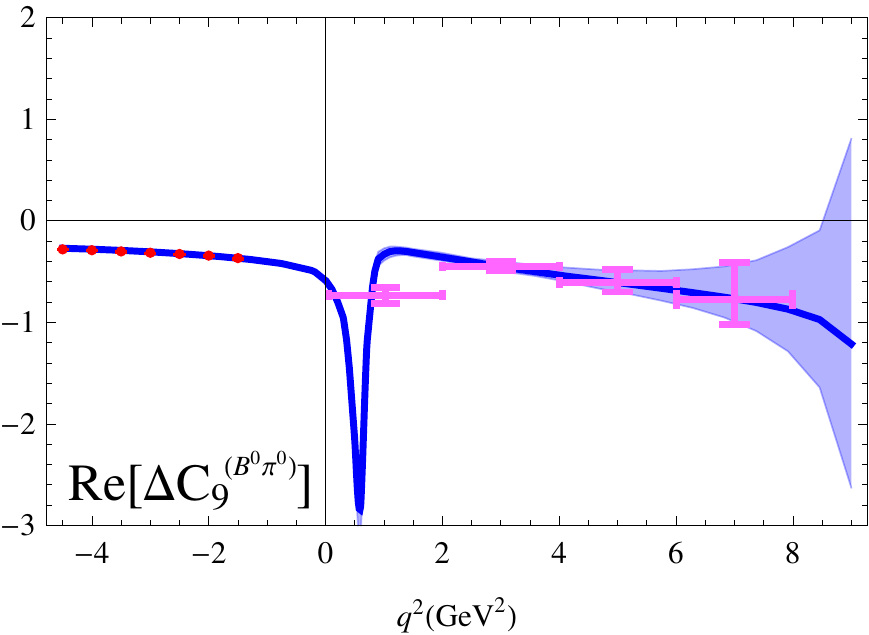}
\includegraphics[scale=0.8]{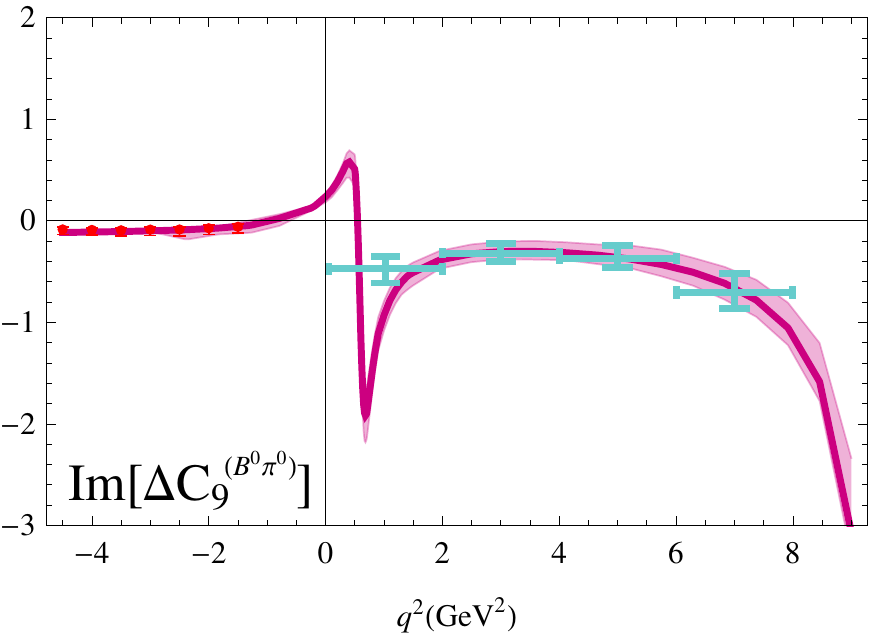}
\caption{\it The same as in fig. \ref{fig:DC9charged}
for $\bar{B^o} \to \pi^0 \ell^+ \ell^-$ (upper panel)
and for $B^o \to \pi^0\ell^+ \ell^-$ (lower panel).
}
\label{fig:DC9neutral}
\end{figure}

Finally, the dispersion relations (\ref{eq:disp1}) take the following form:
\begin{eqnarray}
{\cal H}^{(p)} (q^2)- {\cal H}^{(p)} (q_0^2)
 = 
(q^2 -q^2_0)\Big[\!\!\!\!\!\!\!\!\!\!
\sum\limits_{V=\rho,\omega,J/\psi,\psi(2S)}\!\!\!\!\!\!\!\!\! k_V f_V
\frac{ |A_{BV\pi}^{p}|\exp(i\delta_{BV\pi}^{(p)})}{(m_V^2 - q_0^2 )(m_V^2 -q^2 - i m_V \Gamma_V^{\rm tot})}
\nonumber \\
+\int\limits_{\tilde{s}_0(s_0)}^{4m_D^2} ds 
\frac{\rho_{LO}^{(p)}(s)}{(s-q_0^2)(s-q^2 - i\sqrt{s}\Gamma_{\rm eff}(s))} 
+|a_{p}|\exp(i\phi_a)+
|b_{p}|\exp(i\phi_b)\frac{q^2}{4 m_D^2}
\Big] 
\label{eq:disp2}
\end{eqnarray}
These two relation at $ p=u$ and $p=c$ are then separately 
fitted to the OPE result obtained for the l.h.s. at $q^2<0$.
After that we can use the dispersion form of ${\cal H}^{(p)} (q^2)$
in $q^2>0$ and 
calculate  the correction to the Wilson coefficient 
$\Delta C_9(q^2)$  defined in (\ref{eq:Delta-C9}) in the whole 
large recoil region which we specify as: 
\begin{equation}
4m_\ell^2\leq q^2 \lesssim m_{J/\psi}^2\,.
\label{eq:region}
\end{equation} 
The resulting plots are presented
in Figs.~\ref{fig:DC9charged} and \ref{fig:DC9neutral} for $B^\mp \to \pi^\mp \ell^+ \ell^-$ and 
$B^{\bar{0}0} \to \pi^0 \ell^+ \ell^-$, respectively. 
Instead of showing the fit results for ${\cal H}^{(p)}$ directly, we present the directly 
related, but physically more relevant plots for $\Delta C_9^{(B\pi)}(q^2)$.
At $q^2$  above the $J/\psi$ region  
our approach ceases to work, mostly because the contribution of the 
hadronic dispersion integral (\ref{eq:dispintlargeq2}) to the dispersion relation 
increases and the simple polynomial parametrization cannot be used. This is 
also reflected by the growth of the uncertainties.

 A few comments on the fit procedure of Eq.~(\ref{eq:disp2}) are in order.
Here we only discuss the $B^{\mp} \to \pi^{\mp} +\ell^+ \ell^-$ decay,
as the case of the neutral $B$ decays is very similar.
We represent the results of our calculation at negative $q^2$ values as  ``data" points  and perform a $\chi^2$  fit for our ``model function'' (dispersion relation),
which is valid in both positive and negative $q^2$ regions,  to the 
obtained points. As a technical remark: the fit is performed by 
collecting in the ``data" all parts of Eq.~(\ref{eq:disp2})  which contains 
no fit parameters, i.e. only the resonance contributions (amplitudes and phases) 
and the polynomial continuum parameters are included in the ``model function''.  
Furthermore, we include the error correlation of 
the respective points at negative $q^2$, and, in addition,   
also of  the parameter $f_{B \pi}^{+}(q^2=0)$.  
The errors and the correlation coefficients of the ``data" 
are obtained by varying the input parameters within their intervals 
given in Table~\ref{tab:inputs},
while assuming no correlation between the parameters themselves.
Thus, we include two correlated ``data" sets in the fit for both 
charged $B$-meson transitions, namely, the real 
and imaginary parts  of ${\cal H}^{(u)}$ and ${\cal H}^{(c)}$.
We assume a gaussian error interval of the input parameters for this 
procedure and a maximum error correlation of 80$\%$ for the numerical 
stability, providing also a more conservative estimate.
As expected, the error correlation between the  
``data" points is very large and usually exceeds 80$\%$.
In addition, we find a positive correlation
of, respectively, $\sim 1\%$($10\%$) for the real part and 
of $\sim 40\%$($1\%$) for the 
imaginary part of $H^{(u)}$  ($H^{(c)}$) with  $f_{B \pi}^{+}(q^2=0)$.
The global minima are acceptable with $\chi^2_{min}=1.93$ and 
$\chi^2_{min}=2.53$ for $u$ and $c$, respectively.
The central values quoted here belong to the global minimum, whereas the 68 \% C.L. error 
estimate includes all minima in the $\delta \chi^2<1$ region.  

\section{Observables in $B \to \pi \ell^+ \ell^-$}
\label{sec:obs}
Having calculated the nonlocal amplitudes in a form of the function 
$\Delta C^{(B\pi)}_9 (q^2)$, we substitute this function in the amplitude  
(\ref{eq:Bpill}) of the $B \to \pi \ell^+ \ell^-$ decay 
and predict the observables in the accessible dilepton mass region 
(\ref{eq:region}).

The only element in the complete decay amplitude, 
that was not specified so far, 
is the ratio (\ref{eq:rT}) of tensor and vector $B\to \pi$ form factors
entering the contribution of the $O_{7\gamma}$ operator. 
To obtain it, we evaluate the ratio of LCSR's  for both 
form factors obtained in Ref.~\cite{Duplancic}. The $q^2$-dependence turns
out to be negligible in the whole region of validity of the sum rules,
which covers the region (\ref{eq:region}),    
and we obtain:
\begin{equation}
r_T(q^2)\simeq r_T(0)=0.98 \pm 0.02.
\end{equation}

\begin{figure}[h]\center
\includegraphics[scale=0.8]{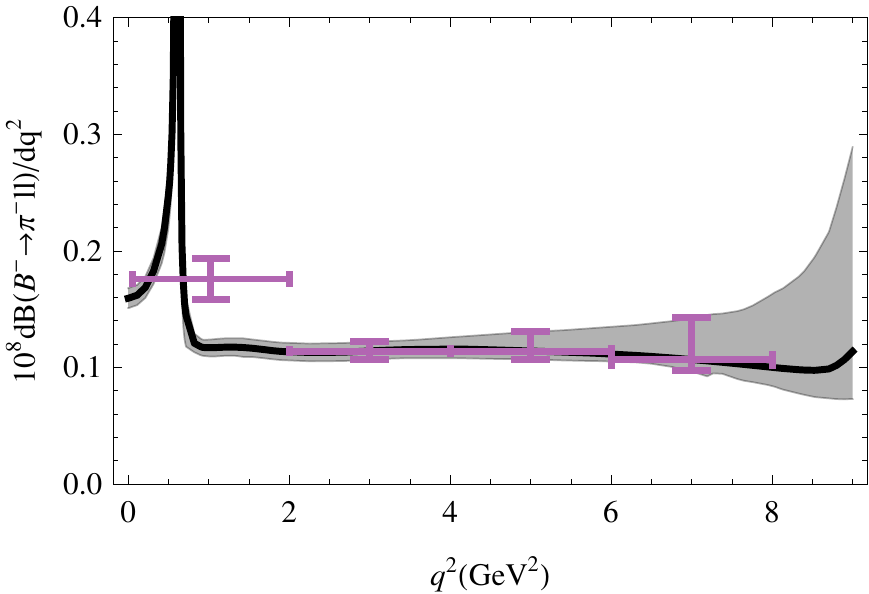}
\includegraphics[scale=0.8]{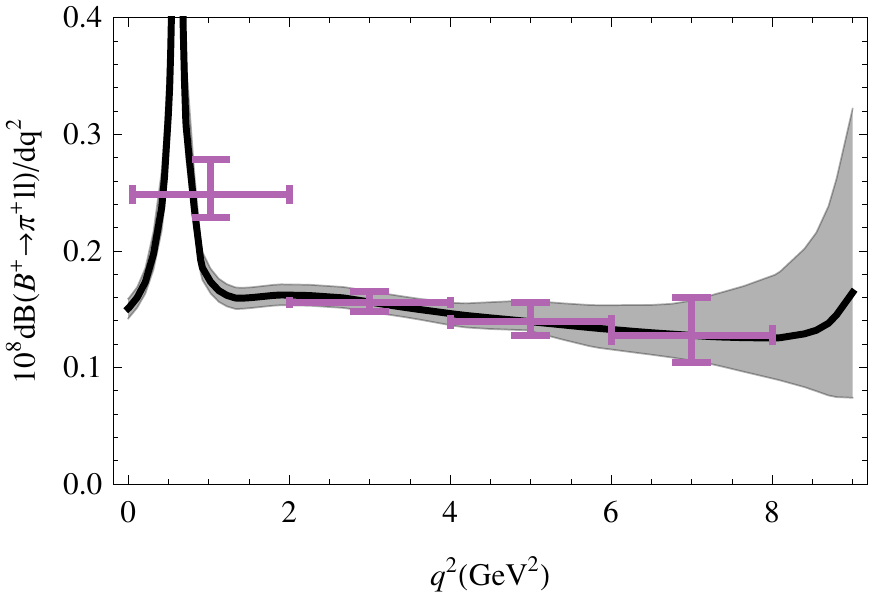}
\caption{\it Dilepton invariant mass spectrum and binned
branching fraction (in GeV$^{\,-2}$) for  
$B^- \to \pi^- \ell^+ \ell^-$ (left panel) 
and $B^+ \to \pi^+ \ell^+ \ell^-$(right panel) 
with 68\% C.L. errors (shaded region and error bars).}
\label{fig:BRBminplus}
\end{figure}

\begin{figure}[h]\center
\includegraphics[scale=0.8]{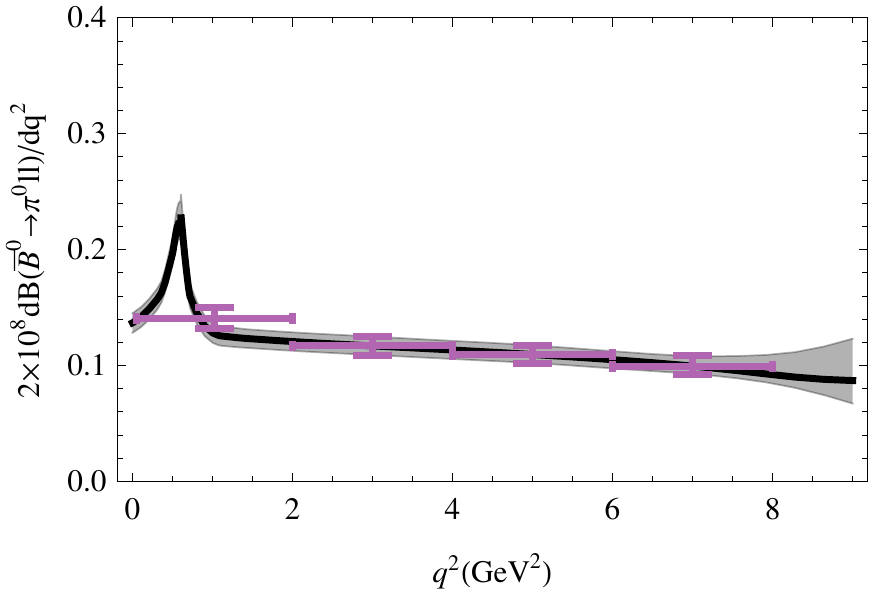}
\includegraphics[scale=0.8]{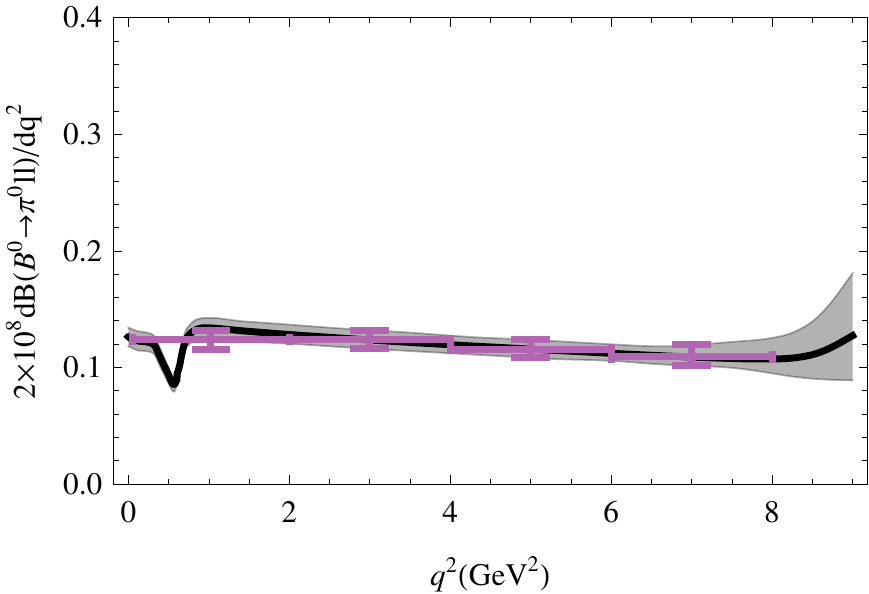}
\caption{\it The same as in fig. \ref{fig:BRBminplus}
for $\bar{B}^0 \to \pi^0 \ell^+ \ell^-$ (left panel) and 
$B^0 \to \pi^+ \ell^+ \ell^-$ (right panel). }
\label{fig:BRBbar00}
\end{figure}

The  observables in  $B \to \pi \ell^+ \ell^-$ include 
the differential branching fraction, direct $CP$-asymmetry 
and isospin asymmetry. 
Note that in SM the angular distribution in $B \to \pi \ell^+ \ell^-$ 
at fixed $q^2$ is reduced to an overall 
factor  $\left(1 - \cos^2 \Theta \right)$ in the double differential distribution
where $\Theta$ is the angle
between the momentum of the lepton $\ell^-$ and the momentum of the $B$-meson
in the dilepton center mass frame. In particular, the 
forward-backward asymmetry in $B \to \pi \ell^+ \ell^-$  vanishes in the SM. 
Hence, it is sufficient to calculate the dilepton invariant mass 
distribution of the branching fraction:
\begin{eqnarray}
\label{eq:dBdqsq}
\frac{1}{\tau_{B^-}}\frac{d B  (B^- \to \pi^- \ell^+ \ell^-)}{d q^2} = 
\frac{G_F^2 \alpha_{\rm em}^2 |\lambda_t|^2}{1536 \pi^5 m_B^3} |f^+_{B\pi}(q^2)|^2  
\lambda^{3/2}(m_B^2,m_\pi^2,q^2)
\nonumber\\
\times \Bigg\{\left| C_9 +\Delta C_9^{B\pi}(q^2) + \frac{2 m_b}{m_B + m_\pi} C_7\,r^T_{B\pi}(q^2)\right|^2
+ \left|C_{10} \right|^2\Bigg\}\,.
\end{eqnarray}
For $\bar{B}^0 \to \pi^0 \ell^+ \ell^-$ the corresonding formula contains
$\tau_{B^0}$ and an additional factor 1/2  reflecting the normalization of the 
$\bar{B}^0\to \pi^0$ form factor. The resulting plots
are presented in Figs.~\ref{fig:BRBminplus}, \ref{fig:BRBbar00}. 
Averaging the above distribution over
$q_1^2\leq q^2\leq q_2^2$  
yields the binned branching fraction defined, e.g., for $B^- \to \pi^- \ell^+ \ell^-$ as: 
\begin{equation}
{\cal B}(B^- \to \pi^- \ell^+ \ell^- [q_1^2,q_2^2])\equiv
\frac{1}{q_2^2-q_1^2}\int\limits_{q_1^2}^{q_2^2}dq^2\frac{d  B  (B^- \to \pi^- \ell^+ \ell^-)}{d q^2}\,.
\label{eq:bin}
\end{equation}
The predicted binned branching fractions within the region (\ref{eq:region}) are 
presented in Table~\ref{Tab:Data-bins} for all 
four flavour/charge combinations.

\begin{figure}[h]\center
\includegraphics[scale=1.0]{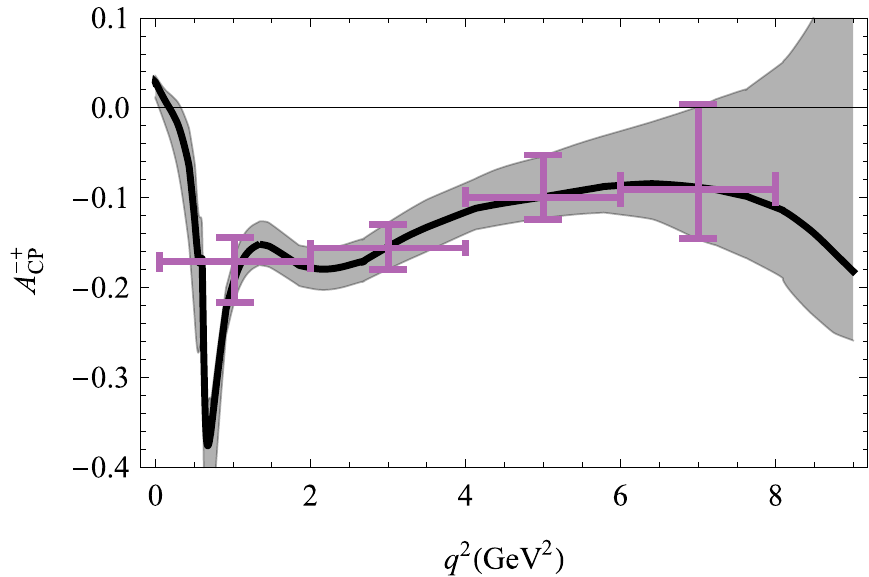}\\
\includegraphics[scale=1.0]{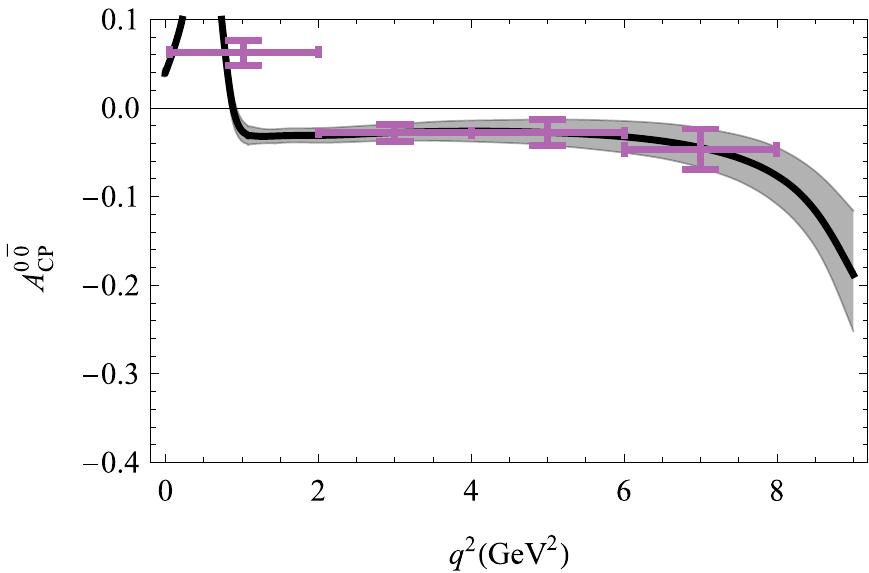}
\caption{\it Direct $CP$-asymmetry in $B^\pm \to \pi^\pm \ell^+ \ell^-$ 
(upper panel) and in $B^{\bar{0}0} \to \pi^0 \ell^+ \ell^-$ (lower panel).}
\label{ACP-B-pi-ell-ell-PM}
\end{figure}

The most interesting characteristics of  the 
$B\to \pi \ell^+ \ell^-$ decay in SM 
is the $q^2$-dependent direct $CP$-asymmetry defined
for the charged $B$-meson modes as:
\begin{eqnarray}
\label{dACP-def-p}
{\cal A}_{CP}^{(-+)} (q^2) = \frac{\displaystyle
d{B}(B^- \to \pi^- \ell^+ \ell^-)/d q^2 
- \displaystyle d{B}(B^+ \to \pi^+ \ell^+ \ell^-)/d q^2
} 
{\displaystyle
d{B}(B^- \to \pi^- \ell^+ \ell^-)/d q^2 
+ \displaystyle d{B}(B^+ \to \pi^+ \ell^+ \ell^-)/d q^2
}
\,.
\end{eqnarray}
The asymmetry for the neutral $B$-meson modes denoted as ${\cal A}_{CP}^{(\bar{0}0)}(q^2)$
has the same expression with $B^-\to \bar{B^0}$, $B^+\to B^0$.
The results obtained for this observable are presented  in Fig.~\ref{ACP-B-pi-ell-ell-PM}.

Anticipating the  future measurements of the $q^2$-averaged bins of 
$CP$-asymmetry, we also calculate  
\begin{equation}
\label{ACP-def-p}
{\cal A}_{CP}^{(-+)}[q_1^2,q_2^2]=
\frac{ {\cal B}(B^- \to \pi^- \ell^+ \ell^- [q_1^2,q_2^2])- 
{\cal B}(B^+ \to \pi^+ \ell^+ \ell^- [q_1^2,q_2^2])}{
{\cal B}(B^- \to \pi^- \ell^+ \ell^- [q_1^2,q_2^2])+ 
{\cal B}(B^+ \to \pi^+ \ell^+ \ell^- [q_1^2,q_2^2])}\,,
\end{equation} 
and the analogous binned asymmetry ${\cal A}_{CP}^{\bar{0}0}[q_1^2,q_2^2]$
for the neutral $B$-meson modes.  Our predictions 
are collected  in Table~\ref{Tab:Data-bins}.
\begin{table}[h]\center
{\small
\begin{tabular}{|c|c|c|c|c|c|}
\hline
Bin [GeV$^2$]& $[0.05,2.0]$& $[2.0,4.0]$& $[4.0,6.0]$& $[6.0,8.0]$&  $[1.0,6.0]$  \\
\hline
${\cal B}(B^-)$ &$0.176^{+0.018}_{-0.018}$&$0.114^{+0.008}_{-0.007}$&$0.114^{+0.016}_{-0.007}$	
&$0.107^{+0.036}_{-0.009}$&$0.126^{+0.013}_{-0.010}$\\
${\cal B}(B^+)$ &$0.249^{+0.030}_{-0.020}$&$0.156^{+0.009}_{-0.008}$&$0.139^{+0.016}_{-0.011}$
&$0.128^{+0.030}_{-0.023}$&$0.168^{+0.016}_{-0.012}$\\
$2\times {\cal B}(\bar{B}^{0})$&	$0.140^{+0.009}_{-0.009}$	&$0.117^{+0.008}_{-0.008}$&	
$0.109^{+0.008}_{-0.008}$	&$0.099^{+0.010}_{-0.007}$&$0.119^{+0.008}_{-0.008}$\\
$2\times {\cal B}(B^0)$&$0.124^{+0.008}_{-0.008}$&$0.124^{+0.008}_{-0.008}$		
&$0.116^{+0.008}_{-0.007}$&$0.109^{+0.011}_{-0.008}$&$0.121^{+0.008}_{-0.008}$\\
\hline
${\cal A}_{CP}^{(-+)}$&$-0.171^{+0.027}_{-0.045}$&$	-0.156^{+0.027}_{-0.024}$&
$-0.099^{+0.047}_{-0.025}$&$-0.091^{+0.093}_{-0.053}$&$-0.143^{+0.035}_{-0.029}$\\
${\cal A}_{CP}^{(\bar 0 0)}$&$\phantom{-}0.063^{+0.014}_{-0.015}$&$-0.028^{+0.010}_{-0.010}$&	
$-0.028^{+0.015}_{-0.015}$&$-0.047^{+0.023}_{-0.023}$&$-0.008^{+0.013}_{-0.013}$\\
${\cal A}_I$ &	$-0.195^{+0.033}_{-0.035}$ & $-0.020^{+0.031}_{-0.032}$ &
$-0.021^{+0.035}_{-0.053}$ & $-0.021^{+0.060}_{-0.100}$&$-0.063^{+0.033}_{-0.040}$\\
\hline
\end{tabular}
}
\caption{\it
Binned branching fractions (in units of 10$^{-8}$ (GeV$^{\, -2}$)), 
direct $CP$-asymmetry  and isospin asymmetry of 
$B\to \pi \ell^+ \ell^-$. }
\label{Tab:Data-bins}
\end{table}

\begin{figure}[h]\center
\includegraphics[scale=1.0]{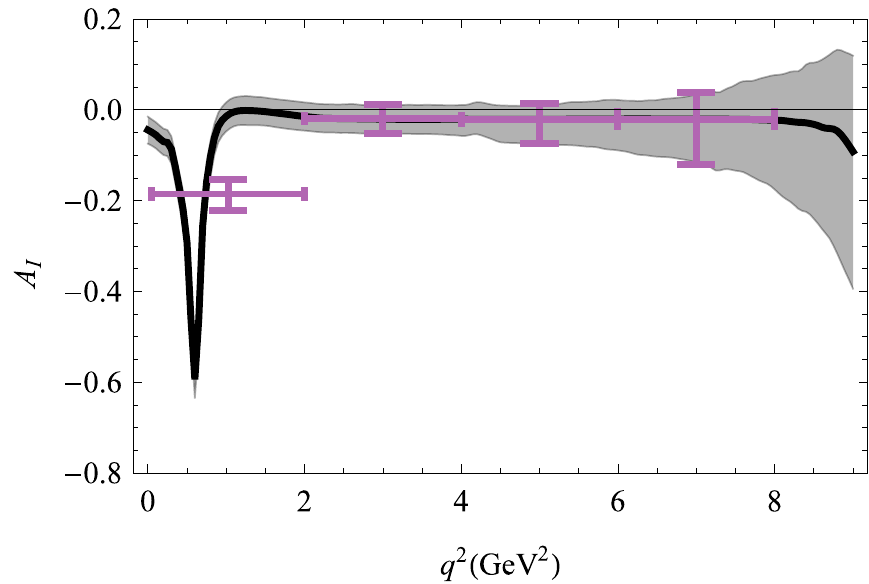}
\caption{\it Differential isospin asymmetry for $B \to \pi \ell^+ \ell^-$ 
decays.}
\label{AI-B-pi-ell-ell}
\end{figure}

Finally, an important indicator of the spectator-dependent 
nonlocal effects, such as weak annihilation, is a nonvanishing  
differential isospin asymmetry 
defined as:
\begin{eqnarray}
\label{dAI-def}
{\cal A}_{I} (q^2) = 
\frac{
2 \displaystyle d\Gamma (\bar B^0 \to \pi^0 \ell^+ \ell^-)/dq^2 
- \displaystyle d\Gamma (B^- \to \pi^- \ell^+ \ell^-)/dq^2
} 
{
2 \displaystyle d\Gamma (\bar B^0 \to \pi^0 \ell^+ \ell^-)/dq^2 + \displaystyle d\Gamma(B^- \to \pi^- \ell^+ \ell^-)/dq^2
}\,,
\end{eqnarray}
where the differential widths are understood as the CP-averaged ones.
Our result is presented in Fig~\ref{AI-B-pi-ell-ell} and 
the corresponding $q^2$-bins of isospin asymmetry:
\begin{equation}
\label{AI-def}
{\cal A}_{I}[q_1^2,q_2^2]=
\frac{ 2\Gamma(\bar{B}^0 \to \pi^0 \ell^+ \ell^- [q_1^2,q_2^2])- 
\Gamma(B^- \to \pi^- \ell^+ \ell^- [q_1^2,q_2^2])}{
2\Gamma(\bar{B}^0 \to \pi^0 \ell^+ \ell^- [q_1^2,q_2^2])+ 
\Gamma(B^- \to \pi^- \ell^+ \ell^- [q_1^2,q_2^2])}\,
\end{equation}
are given in Table~\ref{Tab:Data-bins}.

Concluding the analysis of observables in $B \to \pi \ell^+ \ell^-$, 
we notice that 
the magnitude of the predicted direct $CP$-asymmetry for the charged
$B$ decay modes is quite visible; for the neutral $B$ decays 
this effect is expected to be small. In this and other 
observables our analysis generates
large uncertainties in the region adjacent to $J/\psi$, 
whereas the uncertainties in the $\rho$ and $\omega$ region are 
significantly smaller. This is partly caused by the use of QCDF to fix  
the relative phase between the nonleptonic amplitudes with $\rho$ and $\omega$, 
which probably leads to a slight underestimate
of the errors in the resonance region.

\section{The
$B\to \pi\nu\bar{\nu}$ decay } 
\label{sec:NP}

The semileptonic FCNC decay $B\to \pi\nu\bar{\nu}$  is closely related to the 
charged lepton channel. Theoretically, this process 
is a very clean test of the SM,  involving a single
effective operator similar to $O_{10}$, whereas the nonlocal effects 
studied above  are absent.  Hence, 
we are in a position to predict the branching fraction of this 
decay with a better accuracy than for the $B\to \pi \ell^+\ell^-$.
The only hadronic input in $B\to \pi\nu\bar{\nu}$ is the vector 
$B\to \pi$ form factor. The LCSR form factor \cite{IKMDvD}
given in (\ref{eq:fBpi}), provides an extrapolation 
beyond the large recoil region up to the kinematical limit
$q^2=(m_B-m_\pi)^2$, revealing a good agreement with the 
lattice QCD results in the low recoil region. We use this form factor 
to predict the total branching fraction of  the 
$B\to \pi\nu\bar{\nu}$ decay. 

The effective Hamiltonian encompassing the $b \to d \nu \bar \nu$ transition
in the SM can be written as:
\begin{equation}
{\cal H}_{\rm eff}^{b \to d \nu \bar \nu} =
- \frac{4 G_F}{\sqrt{2}} \lambda_t C_{10\nu} 
\frac{\alpha_{\rm em}}{4 \pi} (\bar d_L \gamma_\mu b_L)
(\bar \nu \gamma^\mu (1 - \gamma_5) \nu)\,,
\label{eq:Onu}
\end{equation}
with the (scale-independent) Wilson coefficient:
\begin{equation}
C_{10\nu} = -\frac{1}{\sin^2 \Theta_W}
\left(X_0 (x_t) + \frac{\alpha_s}{4\pi} X_1 (x_t) \right),
\label{eq:c10nu}
\end{equation}
where $x_t = m_t^2/m_W^2$ and the functions $X_0 (x)$ and $X_1 (x)$ can
be found in Ref.~\cite{BBL_Heff}.

The differential branching fraction
of the $B^- \to \pi^- \nu \bar \nu$ decay summed over neutrino flavours 
has the form:
\begin{eqnarray}
\label{eq:Bpinunu}
\frac{1}{\tau_{B^-}}\frac{d {\cal B} (B^- \to \pi^- \nu \bar \nu)}{dq^2} & \equiv &
\frac{1}{\tau_{B^-}} \sum\limits_{\ell = e, \mu, \tau} \frac{d {\cal B} (B \to \pi \nu_\ell \bar \nu_\ell)}{dq^2} \\
& = & \frac{G_F^2 \alpha_{\rm em}^2}{256 \pi^5 m_B^3} |\lambda_t|^2
|C_{10\nu}|^2 |f^+_{B\pi} (q^2)|^2\lambda^{3/2}(m_B^2,m_\pi^2, q^2)\,.
\nonumber 
\end{eqnarray}
\begin{figure}[t]\center
\includegraphics[scale=0.7]{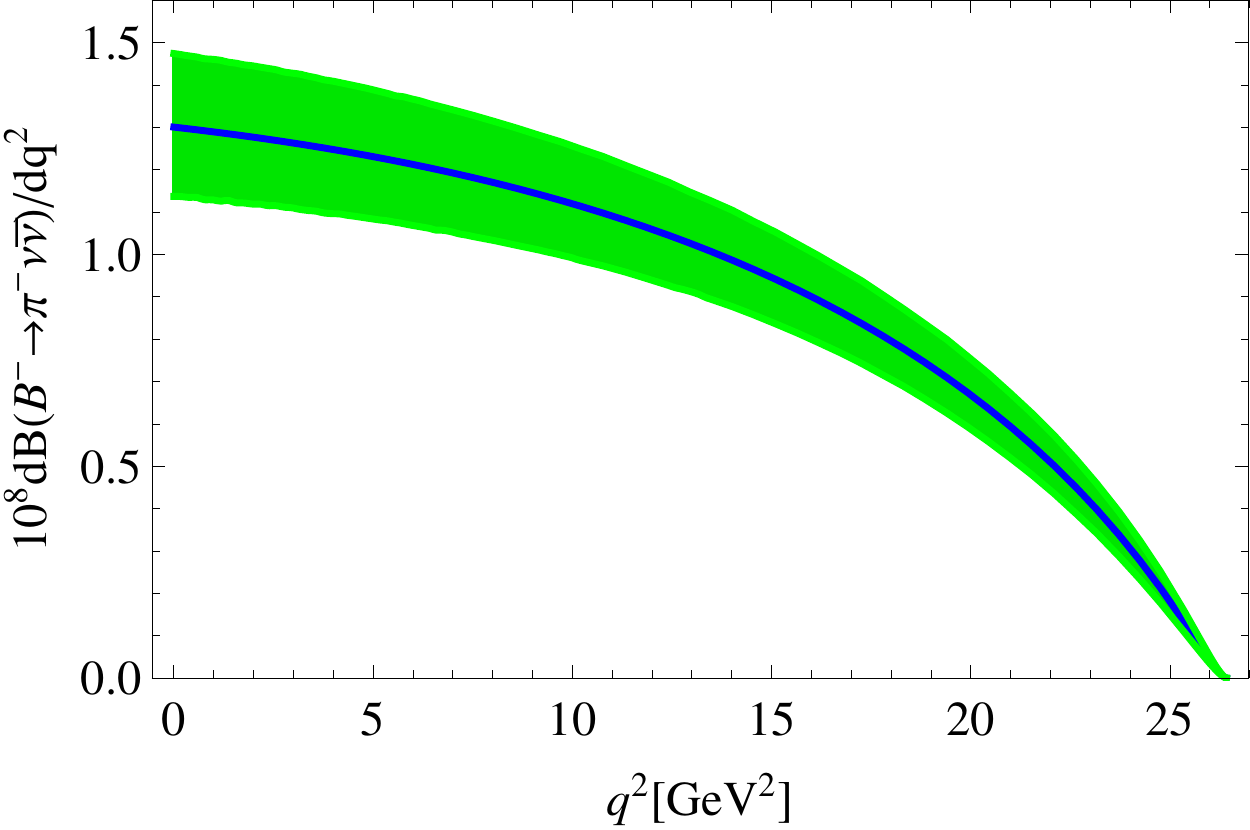}
\caption{The differential branching fraction of the $B^- \to \pi^- \nu \bar \nu$ decay}
\label{fig:Bpinunu}
\end{figure}
Substituting the form factor $f^+_{B\pi} (q^2)$  from  Eq.~(\ref{eq:fBpi}),
the numerical values of the Wilson coefficient $C_{10\nu}=-6.79 $ and other 
parameters in Eq.~(\ref{eq:Bpinunu}), we obtain the differential branching fraction
shown in Fig.~\ref{fig:Bpinunu}. Integrating it over 
$0<q^2<(m_B-m_\pi)^2$ we obtain:
\begin{equation}
{\cal B} (B^- \to \pi^- \nu \bar \nu) = 2{\cal B} (B^0 \to \pi^0 \nu \bar \nu)= (2.39^{+0.30}_{-0.28}) \times 10^{-7}\,.
\end{equation}
Despite the fact that this branching fraction is well within the reach 
of $B$-physics experiments, a severe problem is the identification 
of the final state with respect to the background.

\section{Conclusion}
\label{sec:concl}

In this paper we calculated the hadronic input for the rare FCNC decay $B\to \pi \ell^+\ell^-$ in the large recoil region of the pion, i.e., at  small
and intermediate lepton-pair masses up to the $J/\psi$ mass. 
We focused on the most difficult problem in the theory of these decays: the 
effects generated by a nonlocal overlap 
of the pointlike weak transition with the electromagnetic lepton-pair emission.
At $q^2>0$ the nonlocality involves long distances, including the formation of
hadronic resonances -- the vector mesons. On the other hand, this part
of the decay amplitude  is not simply a background for the FCNC $b\to d \ell^+\ell^-$ 
transition, but provides the strong-interaction phase. The latter,
combined with the CKM phase, generates the unique characteristics of the 
$B\to \pi \ell^+\ell^-$ decay in SM,
that is the   $q^2$-dependent direct CP-asymmetry, suppressed
in the  $b\to s \ell^+\ell^-$ decays.  

To avoid the complications related to the long-distance part of the 
nonlocal effects, we employed the method used earlier 
in Ref.~\cite{KMW_BKll} for $B\to K \ell^+\ell^-$. 
The nonlocal contributions to $B\to \pi \ell^+\ell^-$  transitions 
were calculated one by one,  combining QCDF and LCSRs 
at spacelike $q^2<0$, where the quark-level diagrams are well defined
and the nonlocality is effectively reduced to the distances of $O(1/\sqrt{|q^2|})$.  We also 
used the recently updated \cite{IKMDvD} $B\to \pi$ form factor 
from LCSR. The accuracy of our calculation 
is characterized by taking into account, in addition to the factorizable 
quark-loop effects and the factorizable NLO corrections, also the 
important nonfactorizable contributions: the  soft gluon emission, spectator 
scattering and weak annihilation. 
We then  combined the quark-level calculation with the hadronic dispersion relation 
and fitted the parameters of the latter to access the $q^2>0$ region. The main 
result of our calculation is presented in a form of the  $q^2$-dependent and process-specific 
correction  $\Delta C^{(B\pi)}_9(q^2)$ to the Wilson coefficient 
of the semileptonic operator $O_9$. Apart from the numerical prediction 
for $\Delta C^{(B\pi)}_9(q^2)$, we also  estimated the uncertainties due to 
the input parameter variation.  
We predicted the observables in $B\to \pi \ell^+\ell^-$,
including the differential branching fraction, direct 
$CP$-asymmetry and the isospin asymmetry. The main advantage of the method 
used in this paper is the possibility to access 
the $\rho, \omega$ resonance region and, simultaneously, 
to approach  the charmonium region from below. 

The  accuracy of the calculation carried out  in this paper can be improved further. 
On the theory side  it is worth to calculate the nonlocal contributions 
using entirely LCSRs instead of the  QCDF approximation. This will 
allow one to assess the missing power corrections. Such analysis 
is possible at least for the  weak annihilation and for the hard 
spectator contributions. 
A more elaborated ansatz for the hadronic dispersion relation, 
including the radial excitations of light vector mesons, is also desirable. 
For that, more accurate data on the  $B\to V\pi$ nonleptonic decays 
and a better understanding of the structure of various nonleptonic amplitudes
are needed.

Let us compare our results with the two most recent analyses of the 
$B\to \pi \ell^+\ell^-$ decay. 
In \cite{APR}, only the factorizable nonlocal contributions were taken into account,
approximated by the quark-level diagrams at  positive $q^2$, embedded in 
the short-distance coefficients. Only the 
differential branching fraction was calculated, with no prediction for  the $CP$-asymmetry.
In \cite{HKX}, the QCDF method was systematically used at positive $q^2$, 
therefore the resonance region of $q^2$ was not accessible. 
In the region between 2 GeV$^2$ and 6 GeV$^2$ 
the branching fraction obtained in \cite{HKX} is somewhat smaller than our result, 
whereas the $CP$-asymmetry  is close to our prediction.

We emphasize that our method produces 
a quantitative estimate of the nonlocal effects in the whole
large-recoil region, starting from the kinematical threshold of the lepton-pair production. 
The price to pay is a model dependence of the ansatz for the dispersion relation,
related to the nonleptonic $B\to V\pi$ decays. 
The function $\Delta C^{(B\pi)}_9(q^2)$  obtained in this paper 
can be used in  further analyses of the $B\to \pi \ell^+\ell^-$ decay,
e.g.,  when adding to the decay amplitude in SM certain new physics contributions.
But first of all, it will be very interesting to confront our prediction for the direct  
CP-asymmetry in $B\to \pi \ell^+\ell^-$ with the data. 
Note that the $b\to d \ell^+ \ell^-$ effective  interaction is also probed in 
$B_d\to \mu^+\mu^-$ decay. Its  branching fraction measurement by LHCb and CMS 
collaborations \cite{Bmumu} still leaves 
some room for new physics, making further studies of $b \to d \ell^+ \ell^-$ decays 
very important.

\section*{Acknowledgments}
The work of Ch.H. and  A.K. is supported by 
DFG Research Unit FOR 1873 
``Quark Flavour Physics and Effective Theories'',  Contract No.~KH 205/2-1. 
A.R. acknowledges the support by the Michail-Lomonosov Program of the German
Academic Exchange Service (DAAD) and the Ministry of Education and Science of the
Russian Federation (project No. 11.9197.2014) and the Russian Foundation for
Basic Research (project No. 15-02-06033-a).
We are grateful to
Thorsten Feldmann, Thomas Mannel, Dirk Seidel, Xavier Virto, Danny van Dyk
and Yuming Wang  for useful comments. One of us (A.K.) is grateful to 
Johannes Albrecht  for a helpful discussion.  
    
\section*{Appendix A: Operators and CKM parameters}

In Table \ref{Tab:Wilson-coef-num}  we list the operators entering 
the effective Hamiltonian (\ref{eq:Heff}), and their Wilson coefficients
calculated at LO for three different renormalization scales, 
where $\alpha_{\rm em}=e^2/(4\pi)$ is the electromagnetic coupling, 
$g_s$ is the strong coupling. 
We use the standard conventions for the operators $O_i^{p}$ $(p=u,c)$ 
except the labeling of ${\cal O}_1^{p}$ and ${\cal O}_2^{p}$ is interchanged,
as in \cite{KMW_BKll}.  In the quark-penguin
operators $q=u,d,s,c,b$ and the mass of $d$-quark in 
${\cal O}_{7\gamma}$ and ${\cal O}_{8g}$ is neglected.
The sign conventions for covariant derivatives, $\gamma$-matrices,
left- and right-handed components of the quark fields 
are the same as quoted in the Appendix of \cite{KMW_BKll}. 
\begin{table}[t]\center
\begin{tabular}{|l |l|c|c|c|}
\hline 
Operator &$\mu$ (GeV) &  $2.5$   &    $3.0$   & $4.5$ \\
\hline 
${\cal O}_1^p =  
\left( \bar d_L \gamma_\mu p_L \right) 
\left( \bar p_L \gamma^\mu b_L \right)$ &$C_1$  & 1.169  & 1.148 & 1.111 \\         
\hline
${\cal O}_2^p  =  
\left( \bar d_L^i \gamma_\mu p_L^j \right) 
\left( \bar p_L^j \gamma^\mu b_L^i \right)$ &$C_2$  & -0.360   & -0.324 & -0.255 \\         
\hline
${\cal O}_3  =  
\left( \bar d_L \gamma_\mu b_L \right) 
\sum\limits_q \left ( \bar q_L \gamma^\mu q_L \right) $ &$C_3 (\times 10^{-2})$  & 1.700  & 1.503 & 1.144 \\         
\hline
${\cal O}_4  = 
\left( \bar d_L^i \gamma_\mu b_L^j \right) 
\sum\limits_q \left ( \bar q_L^j \gamma^\mu q_L^i \right)$  &$C_4 (\times 10^{-2})$  & -3.602  & -3.271 & -2.630 \\         
\hline
${\cal O}_5  =  
\left( \bar d_L \gamma_\mu b_L \right) 
\sum\limits_q \left ( \bar q_R \gamma^\mu q_R \right)$&$C_5 (\times 10^{-2})$  & 0.985  & 0.910 & 0.756 \\         
\hline
${\cal O}_6  =  
\left( \bar d_L^i \gamma_\mu b_L^j \right) 
\sum\limits_q \left ( \bar q_R^j \gamma^\mu q_R^i \right)$&$C_6 (\times 10^{-2})$  & -4.829  & -4.258 & -3.236 \\         
\hline
${\cal O}_{7\gamma}  =  - \frac{e \, m_b}{16 \pi^2}   
\left ( \bar d_L \sigma^{\mu \nu} b_R \right ) F_{\mu \nu}$&$C_7^{\rm eff}$  & -0.356  & -0.343 & -0.316 \\         
\hline
${\cal O}_{8g}  =  - \frac{g_s m_b}{16 \pi^2}   
\left ( \bar d_L^i \sigma_{\mu \nu} (T^a)^{i j} b^j_R \right ) G^{a\, \mu \nu}$&$C_8^{\rm eff}$  & -0.166  & -0.160 & -0.150 \\         
\hline
${\cal O}_9 =  \frac{\alpha_{\rm em}}{4 \pi}   
\left ( \bar d_L \gamma^\mu b_L \right ) 
\left ( \bar \ell \gamma_\mu \ell \right )$
&$C_9$  & 4.514  & 4.462 & 4.293 \\         
\hline
${\cal O}_{10}  =  \frac{\alpha_{\rm em}}{4 \pi}   
\left ( \bar d_L \gamma^\mu b_L \right) 
\left ( \bar \ell \gamma_\mu \gamma_5 \ell \right )$&$C_{10}$  & -4.493  & -4.493 & -4.493 \\         
\hline
\end{tabular}
\caption{\it Effective operators  and Wilson coefficients.}
\label{Tab:Wilson-coef-num}
\end{table}
The electroweak parameters used to calculate the coefficients 
$C_i$ are \cite{PDG14}
\begin{eqnarray}
 \alpha_{\rm em} = \displaystyle\frac{1}{129}\,,~~ 
\sin^2 (\Theta_W) = 0.23126\,,~~m_W= 80.385 \mbox{ GeV}\,, 
\nonumber \\
G_F = 1.1663787 \times 10^{-5} \mbox{ GeV}^{-2}\,, 
~m_z = 91.186 \mbox{ GeV}\,,~~
m_t= 173.3 \mbox{ GeV}\,.
\end{eqnarray}
We use the CKM mixing matrix  in term of Wolfenstein parameters 
$$ 
V_{\rm CKM} = 
\left(
\begin{array}{ccc}
1 - \lambda^2/2               & \lambda              & A \lambda^3 (\rho - i \eta) \\
- \lambda                     & 1-\lambda^2/2        & A \lambda^2                 \\
A \lambda^3 (1 -\rho -i \eta) & -A \lambda^2         & 1                           \\
\end{array}
\right), 
$$
taking into account that $\rho \simeq \bar \rho \left(1 + \lambda^2/2\right)$
and  $\eta \simeq \bar \eta \left(1 + \lambda^2/2\right)$ and 
using the current values \cite{PDG14} 
obtained from the global CKM fit:
\begin{eqnarray}
\lambda = 0.22537 \pm 0.00061, & \quad & A = 0.814^{+0.023}_{-0.024},
\nonumber \\
\bar \rho = 0.117 \pm 0.021, & \quad & \bar \eta = 0.353 \pm 0.013.
\end{eqnarray}
This results in the following combinations of CKM elements we use:
\begin{eqnarray}
\lambda_u/\lambda_t= - 0.0274 - i \, 0.3896,  \quad |\lambda_u/\lambda_t| = 0.3906, 
\quad {\rm arg}(\lambda_u/\lambda_t) = -94.02^\circ
\nonumber \\
\lambda_c/\lambda_t= - 0.9719 + i \, 0.3998,  
\quad |\lambda_c/\lambda_t| = 1.0509, 
\quad {\rm arg}(\lambda_c/\lambda_t) = 157.64^\circ\,.
\end{eqnarray}

\section*{Appendix B: Amplitudes of 
$B\to \rho(\omega)\pi$  in QCDF}
Here we present the expressions of the QCDF amplitudes \cite{BN03} for 
the $B^-\to(\rho^0,\omega)\pi^-$ nonleptonic decays.
Our  operators differs from the ones in \cite{BN03} by a  factor 1/4
whereas the labeling of $O_{1,2}$ is the same.
The expressions for the parts of $B^- \to \rho^0 \pi^-$ and 
$B^- \to \omega \pi^-$ amplitudes multiplying $\lambda^p$ ($p=u,c)$ are:
\begin{eqnarray}
A_{B^-\rho\pi^-}^p  =  
A_{\pi \rho} \bigg(\delta_{pu}\big[\alpha_2 (\pi \rho) - \beta_2 (\pi \rho)\big] 
- \alpha_4^p (\pi\rho) -\beta_3^p (\pi\rho) \bigg)
\nonumber \\
+  A_{\rho\pi} \bigg(\delta_{pu} \big[\alpha_1 (\rho \pi) + \beta_2 (\rho \pi)\big]
+\alpha_4^p (\rho\pi) + \beta_3^p (\rho\pi)\bigg),
\label{Ampl-Bm-to-rho0-pim}   
\end{eqnarray}
\begin{eqnarray}
A_{B^-\omega\pi^-}^p = 
A_{\pi \omega} \bigg(\delta_{pu}\big[\alpha_2 (\pi \omega) + \beta_2 (\pi \omega)\big] 
+ 2 \alpha_3^p (\pi \omega) + \alpha_4^p (\pi\omega) +\beta_3^p (\pi\omega)\bigg)
\nonumber \\
+ A_{\omega \pi} \bigg(\delta_{pu} \big[\alpha_1 (\omega \pi) + \beta_2 (\omega \pi)\big]
+\alpha_4^p (\omega\pi) + \beta_3^p (\omega\pi) \bigg),
\label{Ampl-Bm-to-omega-pim}
\end{eqnarray}
where the factorized combinations of form factors and decay constants are:
\begin{equation}
A_{\pi \rho(\omega)} = \frac1{2\sqrt{2}}  f^+_{B\pi} (m_\rho^2) f_{\rho(\omega)},~~~
A_{\rho(\omega) \pi} =  \frac{1}{2\sqrt{2}}  A^0_{B \rho(\omega)} (0) f_\pi\,.
\end{equation}
In addition to the already introduced notation,  
$A^0_{B \rho(\omega)}(0)$ in the above is the relevant $B\to \rho(\omega)$ 
form factor taken at $q^2=0$, neglecting the pion mass squared; 
for the $B\to \pi$ form factor we  approximate $m_\rho^2=m^2_\omega$. 
The parameters $\alpha_i^p (M_1 M_2)$ are defined 
as follows \cite{BN03}: 
\begin{equation}
\alpha_i (M_1 M_2) = a_i (M_1 M_2), \quad i = 1,2,
\end{equation}
\begin{equation}
\alpha_3^p = 
\left\{
\begin{array}{ll}
a_3^p (M_1 M_2) + a_5^p (M_1 M_2), & \, {\rm if} \,  M_2 = \rho, \omega,  \\
a_3^p (M_1 M_2) - a_5^p (M_1 M_2), & \, {\rm if} \,  M_2 = \pi,  \\
\end{array} 
\right. 
\end{equation}
\begin{equation}
\alpha_4^p = 
\left\{
\begin{array}{ll}
a_4^p (M_1 M_2) + r_\chi^{M_2} a_6^p (M_1 M_2), & \, {\rm if} \,  M_2 = \rho, \omega,  \\
a_4^p (M_1 M_2) - r_\chi^{M_2} a_6^p (M_1 M_2), & \, {\rm if} \,  M_2 = \pi,  \\
\end{array} 
\right. 
\end{equation}
where
\begin{equation}
r_\chi^\pi = \frac{2 m_\pi^2}{m_b \,(m_u+m_d)}, \quad 
r_\chi^{\rho, \omega} = \frac{2 m_{\rho,\omega}}{m_b}\frac{f_{\rho,\omega}^\perp}{f_{\rho,\omega}},
\end{equation}
and $f_{\rho(\omega)}^\perp$ is  the vector-meson transverse decay
constant, defined as 
\begin{equation}
\label{eq:fT}
\langle 0 | \bar{q}\sigma^{\mu\nu}q | V (q) \rangle = 
ik^{\perp}(\varepsilon_{V}^\mu q^\nu-\varepsilon_{V}^\nu q^\mu) f_{V}^\perp\,
\end{equation}
with $k^{\perp}=1/\sqrt{2}$ for $q=u$, $V=\rho^0,\omega$. 

The quantities $a_i^p (M_1 M_2)$ have the form \cite{BBNS}:
\begin{eqnarray}
a_i^p (M_1 M_2) & = & \left(C_i + \frac{C_{i\pm 1}}{N_c} \right) N_i (M_2) 
\nonumber \\
& + & \frac{C_{i \pm 1}}{N_c} \frac{C_F \alpha_s}{4 \pi}
\left[V_i (M_2) + \frac{4 \pi^2}{N_c} H_i (M_1 M_2) \right] + P_i^p (M_2)\,,
\label{eq:qcdf}
\end{eqnarray}
where the upper (lower) signs apply when $i$ is odd (even);
$N_i (M_2) = 0 \,$ for $i = 6$ and $M_2 = \rho,\omega$ and $N_i (M_2) = 1$
in all other cases.
The parameters $\beta_i^p (M_1 M_2)$ involve the weak annihilation 
contributions:
\begin{equation}
\beta_i^p (M_1 M_2) \equiv \frac{-f_B f_{M_1} f_{M_2}}{2\sqrt{2}m_B^2A_{M_1 M_2}} b_i^p (M_1 M_2)\,.
\label{eq:WAbeta}
\end{equation}

The expressions used for the separate contributions in Eqs. 
(\ref{eq:qcdf}), (\ref{eq:WAbeta}):  $V_i (M_2)$ (one-loop vertex
correction), $H_i (M_1 M_2)$ (hard-spectator scattering), $P_i^p (M_1 M_2)$
(penguin contractions) and  $b_i^{(p)} (M_1 M_2)$ (weak annihilation) 
can be  found in  \cite{BBNS,BN03}. They were
calculated in  QCDF in terms of the perturbative kernels convoluted
with the DAs of the $B$ meson, pion and $\rho$($\omega)$ meson. The latter DAs 
include the Gegenbauer moments $a_2^{\rho(\omega)}$ and 
$a_2^{\rho, \perp}$, similar to the ones that are contained in the pion
twist-2 DA (\ref{eq:varphi}). 

For the numerical analysis of the  $B^-\to \rho(\omega) \pi$ 
amplitudes we need additional input parameters listed 
in Table~\ref{tab:input2}, where, in order to decrease the uncertainty,  
the  $A_0^{B\omega}(0)$  form factor is calculated multiplying the 
ratio $A_0^{B\omega}(0)/f_+^{(B \pi)}(0)$ 
obtained from the LCSRs with the $B$-meson DAs \cite{KMO}
with  the  form factor $f_+^{(B \pi)}(0)$ taken from the 
most accurate LCSR with pion DAs \cite{IKMDvD}.

\begin{table}[h]\center
\begin{tabular}{|l|c|}
\hline
Parameter & Ref.\\
\hline
$(m_u+m_d)( 1 \mbox{GeV})= 7.0 ^{+1.4}_{-0.4} $ MeV &\cite{PDG14}\\
\hline
$f_+^{(B \pi)} (m_\rho^2) = 0.316 \pm 0.021 \hspace*{20mm}$ &\cite{IKMDvD}\\
\hline
$A_0^{B\rho} (0) = 0.396^{+0.039}_{-0.031}$ &\cite{KMO}\\
$A_0^{B\omega}(0) \simeq A_0^{B\rho} (0)$ &\\
\hline
$f_\rho^{\perp}( 1 \mbox{GeV})  = (0.160 \pm 0.010) $ GeV& \cite{BZ04}\\
$f_\omega^{\perp}( 1 \mbox{GeV}) = (0.145 \pm 0.010) $ GeV &\\
$a_2^{\rho,\omega}( 1 \mbox{GeV}) = a_2^{\rho,\omega \perp} ( 1 \mbox{GeV})= 0.09^{+0.10}_{-0.07} $ &\\
\hline
\end{tabular}
\label{tab:input2}
\caption{\it Additional input parameters related to the light vector mesons 
and used in the QCDF amplitudes of $B \to \rho(\omega) \pi$  decays.}
\end{table}


\end{document}